\documentclass[preprint,3p,times,twocolumn, nonatbib]{elsarticle}
\usepackage{rotating,color,subfigure,amssymb}
\usepackage{alphalph}
\usepackage[T1]{fontenc}
\usepackage{float}
\journal{Nuclear Instruments and Methods in Physics Research Section A}
\usepackage[utf8]{inputenc}
\usepackage[sorting=none]{biblatex}
\usepackage{notoccite}
\usepackage{physics}
\usepackage{mathrsfs}
\usepackage{amsmath}
\usepackage{amssymb}
\usepackage{xcolor}
\usepackage{stfloats}
\usepackage{graphicx}% Include figure files
\usepackage{dcolumn}% Align table columns on decimal point
\usepackage{bm}% bold math
\usepackage{mathtools,slashed}
\bibliography{references.bib}
\begin{document}

%\preprint{APS/123-QED}

\title{Developing Multi-Dimensional, Sequential sWeighting Routines for Reaction Selection with Emphasis on Kaon Identification at CLAS12}% Force line breaks with \\

\author[UoY]{A. Acar}
\author[UoY]{M.~Bashkanov}\ead{mikhail.bashkanov@york.ac.uk}
\author[UoY]{D.P. Watts}
\author[UoY]{M. Nicol}

\address[UoY]{School of Physics, Engineering and Technology, Department of Physics, University of York, Heslington, York, Y010 5DD, UK}
%\email{mikhail.bashkanov@york.ac.uk}

\date{\today}% It is always \today, today,

\begin{abstract}
In nuclear and particle physics experiments, event selection is often a multidimensional classification problem which involves several correlated observables. The widely used sWeight technique provides a statistically rigorous means of separating signal and background using discriminating variables, but its conventional formulation is generally applied to a single discriminating variable or a simultaneous multidimensional fit. In this paper we present a novel extension of the widely used sWeight technique, in which an arbitrary number of discriminating variables can be used to construct a  multidimensional sequential framework. This method allows the extraction of signal distributions in controlled variables in a rigorous way. Particular attention is given to the treatment of signal and background correlations, as well as the propagation of statistical uncertainties through successive weighting stages. The method yields high-purity signal samples even in the presence of large and correlated backgrounds. As a benchmark application, the validity of this technique is demonstrated on kaon identification in a dataset collected with the CLAS12 detector at the Thomas Jefferson National Laboratory for cascade baryon searches. The final workflow is shown to be robust and consistent across different run periods and beam/detector conditions, demonstrating minimal systematical uncertainties associated with the method. Statistical uncertainties are proved to be evaluable using a bootstrap-based procedure. The method provides a general framework for multidimensional event weighting and is intended for application in complex analyses requiring robust signal selection in the presence of complex correlated backgrounds.
\end{abstract}

\maketitle

%\tableofcontents
\section{\label{sec:Intro0} Introduction}

Separation of signal and background contributions is a central challenge in experimental physics analyses. In the simplest case, a single discriminating observable is used to separate signal and background components, while a second variable of physical interest is treated as a control quantity. For this class of problems, a mathematically rigorous procedure called sWeights has been developed~\cite{splots} and has seen widespread application in nuclear and particle physics in recent years~\cite{Aaij_2015, Hayrapetyan_2026, Ablikim_2025}. In realistic analyses, however, this simplified scenario is rarely applicable. Instead, event classification typically relies on multiple correlated discriminating observables, and the separation of signal and background must be performed in a multidimensional space. This introduces additional complications related to correlations and the propagation of statistical uncertainties across stages of the analysis.

In this paper, we extend the sWeight formalism from the conventional one-dimensional case to a multidimensional framework, with a sequential approach. We examine under which conditions the multidimensional problem can be factorised, and we quantify the limitations of such an approach in the presence of intrinsic correlations between signal and background distributions. In addition, we present a method to evaluate statistical uncertainties in this more general weighting procedure. The approach is validated using data from the CLAS12 Collaboration experiment at Jefferson Lab, applied to the reaction $e^-p\to e^- K^+K^+ X$, which is characterised by low signal to noise ratio and strongly correlated background contributions.

The paper is organised as follows: Section~\ref{sec:IntroCLAS12} and~\ref{sec:PID} describe CLAS12 experiment  and associated particle identification framework. This is followed by an introduction to the sWeighting method and a discussion of its limitations in Sections~\ref{sec::SWeights} and~\ref{sec::1Dissues}, respectively. In Section~\ref{sec::2D}, we present the extension of the sWeight formalism to a two-dimensional case, including the parametrisation of signal and background components. Section~\ref{chap:seq} then describes the sequential application of the multidimensional sWeighting procedure for kaon identification. Section~\ref{chap::PiPr} discusses the importance of an appropriate signal and background parametrisation, and demonstrates a data-driven approach for extracting background shapes. The evaluation of statistical uncertainties and their propagation is presented in Section~\ref{sec::Boot}. Finally, Section~\ref{sec::corr} reports Monte Carlo studies of the method’s performance as a function of signal–background correlation strength, providing mathematical validation of the approach.

\section{\label{sec:IntroCLAS12} CLAS12}
The CEBAF Large Acceptance Spectrometer (CLAS12)~\cite{CLAS12}, located in Hall B at Jefferson Lab, is designed for the detection of charged and neutral particles over a large solid-angle acceptance approaching $4\pi$, while operating at high luminosities. The spectrometer provides high-resolution charged-particle tracking, enabled by its configuration around a large toroidal magnetic field.

\begin{figure}[h!] 
\centering
\includegraphics[width=0.9\linewidth] 
{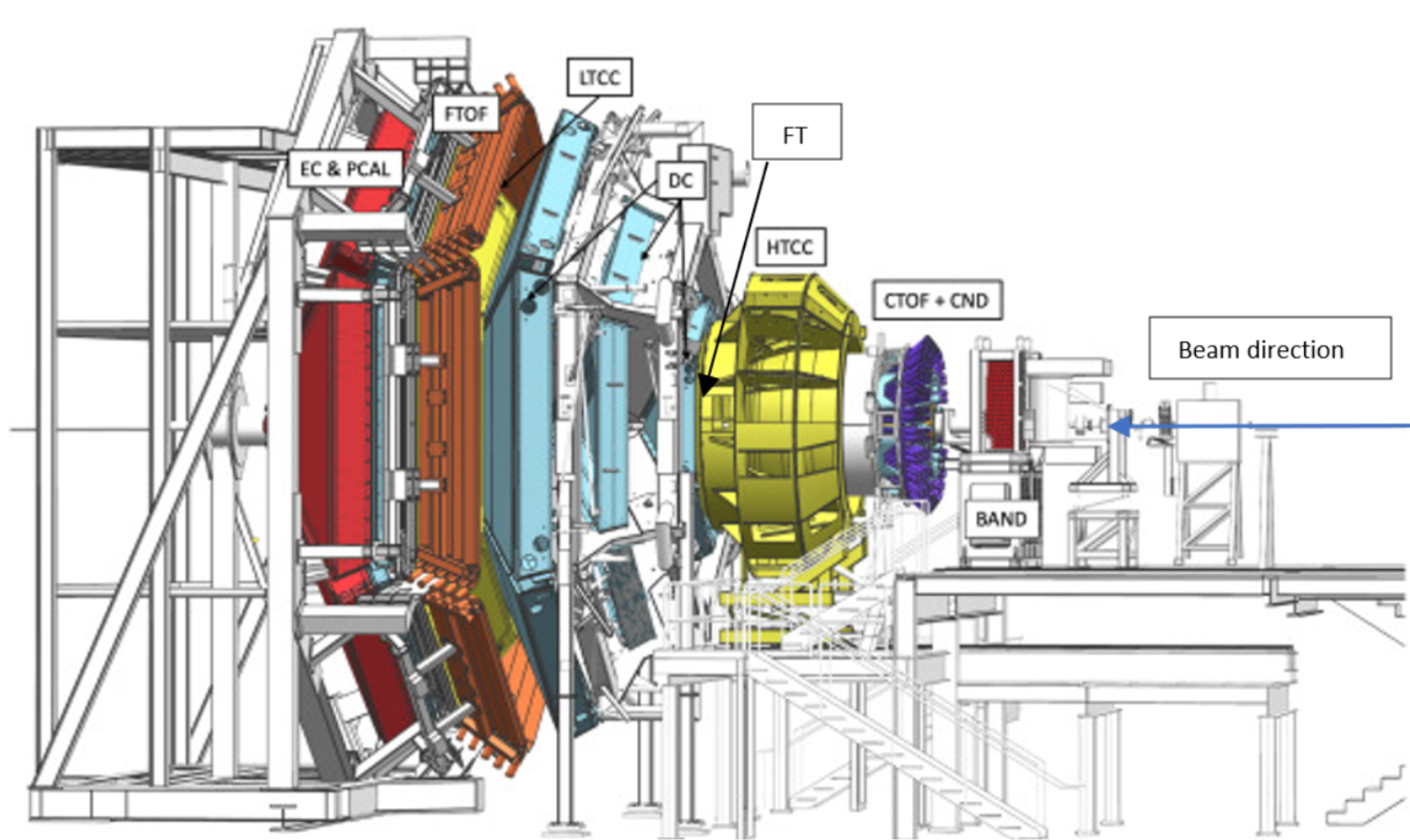}
\caption{The rendering of CLAS12, with important components of the spectrometer labelled.}
\label{fig: clas12}
\end{figure}

The rendering of the detector is shown in Figure~\ref{fig: clas12}. CLAS12 consists of three main sections: Forward Detector (FD), Central Detector (CD) and Forward Tagger (FT). The Central Detector (CD) covers angular range $35^{\circ}<\theta<125^{\circ}$. The CD has its own particle identification and tracking detectors: the Central Vertex Tracker (CVT) \cite{SVT}, Central Time of Flight (CTOF)~\cite{CTOF}, Central Neutron Detector (CND)~\cite{CND} and Back Angle Neutron Detector (BAND)~\cite{BAND}. The CVT combines silicon and micromegas detectors to determine the vertex of charged particles that scatter from the target, and measures momentum by mapping the curvature of tracks in the magnetic field of the solenoid. The CTOF provides time-of-flight measurements that are used to identify charged particles. The CND detects neutrons that have momenta between 0.2 and 1 GeV. BAND is used to detect backward neutrons. 

The Forward Tagger~\cite{FT}, designed and constructed by INFN Genoa, CEA-Saclay and the University of York, covers the polar angular range $2.5^{\circ}<\theta<4.5^{\circ}$, giving us the opportunity to explore events in which we have quasi-real photoproduction. It is composed of a calorimeter, a micro-strip gas tracker and a hodoscope. The calorimeter is used to measure the deposited energy of the electromagnetic shower. The micro-strip gas tracker is tasked with measuring the scattering angles of the charged particles, and the hodoscope distinguishes electrons from high energy photons. 

The FD covers the angular range $5^{\circ}<\theta<45^{\circ}$. The FD’s subsystems consist of:

\begin{itemize}
    \item Drift Chambers (DC)
    \item Cherenkov Detectors: Low Threshold Cherenkov Counter (LTCC), High Threshold Cherenkov Counter (HTCC), and Ring Imaging Cherenkov Detector (RICH)
    \item Forward Time Of Flight (FTOF)
    \item Electromagnetic Calorimeters (ECAL) and Pre-shower Calorimeter (PCAL)
\end{itemize}
The toroidal magnetic field of the spectrometer can be operated in two polarity configurations, commonly referred to as \emph{inbending} and \emph{outbending} modes. In the \emph{inbending} mode, negatively charged particles, such as electrons, are bent towards the beamline. In the \emph{outbending} mode, the field polarity is reversed, and negatively charged particles are bent away from the beamline.
The DC \cite{DC} consists of six sectors, each containing 36 sense-wire layers arranged into three regions. These regions are positioned before, within, and after the toroidal magnetic field, providing redundant tracking measurements for charged-particle trajectory reconstruction. For more details about the CLAS12 detector systems see Ref.~\cite{CLAS12}.

The data that will be used in this work are part of Run Group A (RGA) beamtime, taken during Fall 2018 and Spring 2019 by the CLAS12 detector, utilizing a 5 cm unpolarized liquid-hydrogen target. Data taken with the torus magnet in \emph{outbending} and \emph{inbending} modes are available for Fall 2018, whereas only \emph{inbending} data are available for Spring 2019. During RGA, the electron beam was longitudinally polarized with an energy of 10.6 GeV during Fall 2018, and 10.2 GeV during Spring 2019. 

\section{\label{sec:PID} Method of Particle Identification}
There are many methods to separate different species of particles at detector level. One of these methods uses the relativistic speed of the particle, $\beta$: 
\begin{align}
\beta = v/c
\end{align}
where $v$ is the speed of the particle and $c$ is the speed of light in vacuum, together with the momentum of the particle, to calculate its mass. In order to obtain $v$, we need information about time ($t$) and distance ($d$):
\begin{align}
v = \frac{d}{t}
\end{align}

In Hall B, the electron beam comes in bunches of either 2 or 4 ns. An electron or other charged particle, assuming they have speed $c$, in the event determines the \emph{start time}. The beam bunch information is used to improve the resolution of the event start time. The FTOF system~\cite{FTOF} is responsible for providing the \emph{stop time} information to distinguish different species of particles using plastic scintillators.

\begin{figure} [!h]
\centering
\includegraphics[width=0.9\linewidth]
{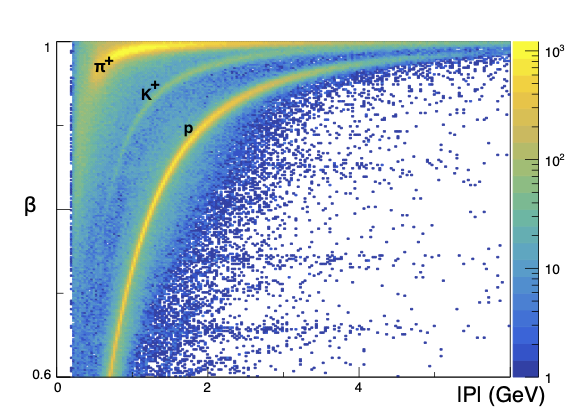}
\caption{Data taken in Fall 2018 with the \emph{inbending} torus magnet configuration. $\beta$ vs magnitude of 3-momentum, $|P|$, for $\pi^+, K^+$, and p.(Note the log-Z scale). }
\label{fig: BetaBefore}
\end{figure}

In this particular case with CLAS12, electrons arrive at the target in narrow $\sim 150$~fs bunches every 4 nanoseconds, hence the production time of a particle is known with very high precision. By mapping the trajectory, one can obtain the distance between production, \emph{start point}, and the \emph{stop point} at the ToF wall.
At the end of each reconstructed track, the ToF detectors measure the particle arrival time~\cite{FTOF,DC}. Using the curvature inside the toroidal magnetic field of the FD torus magnet, $\beta$ can be calculated, and a momentum vector can be extracted from the fitted track. The resulting $\beta$ versus momentum distribution for positively charged tracks is shown in Fig.~\ref{fig: BetaBefore}, exhibiting distinct bands corresponding to pions, kaons, and protons. The kaon band is significantly suppressed compared to the neighboring pion and proton bands (note the log-Z scale on the plot), reflecting the relatively low kaon yield compared to non-strange particles~\footnote{For this energy range production of any additional kaon leads to about 3 orders of magnitude suppression in event yield~\cite{wreo32365}}. This imbalance makes kaon identification particularly sensitive to background contamination and therefore necessitates more sophisticated background suppression techniques.
As momentum increases, the particle bands converge, resulting in reduced separation and increased contamination from background in the high-momentum region. In addition to the in-time particle bands, Fig.~\ref{fig: BetaBefore} exhibits faint horizontal bands. These originate from particles produced in neighboring electron bunches for which an incorrect event start time was assigned. It is convenient to study mass-momentum, $m-p$, distributions rather than velocity-momentum, $\beta-p$, for particle identification, as the out of time background is more readily modeled in this space.
One can calculate the mass of a particle from its given $\beta$ and $p$:
\begin{align} 
m^2 = \frac{p^2}{\beta^2} - p^2,
\label{kmass}
\end{align}
where $m$ is the measured mass of the particle, $\beta$ is the dimensionless relativistic velocity and $p$ is the 3-momentum.

\begin{figure}[!h]
\begin{center}
\includegraphics[width=0.5\textwidth,angle=0]{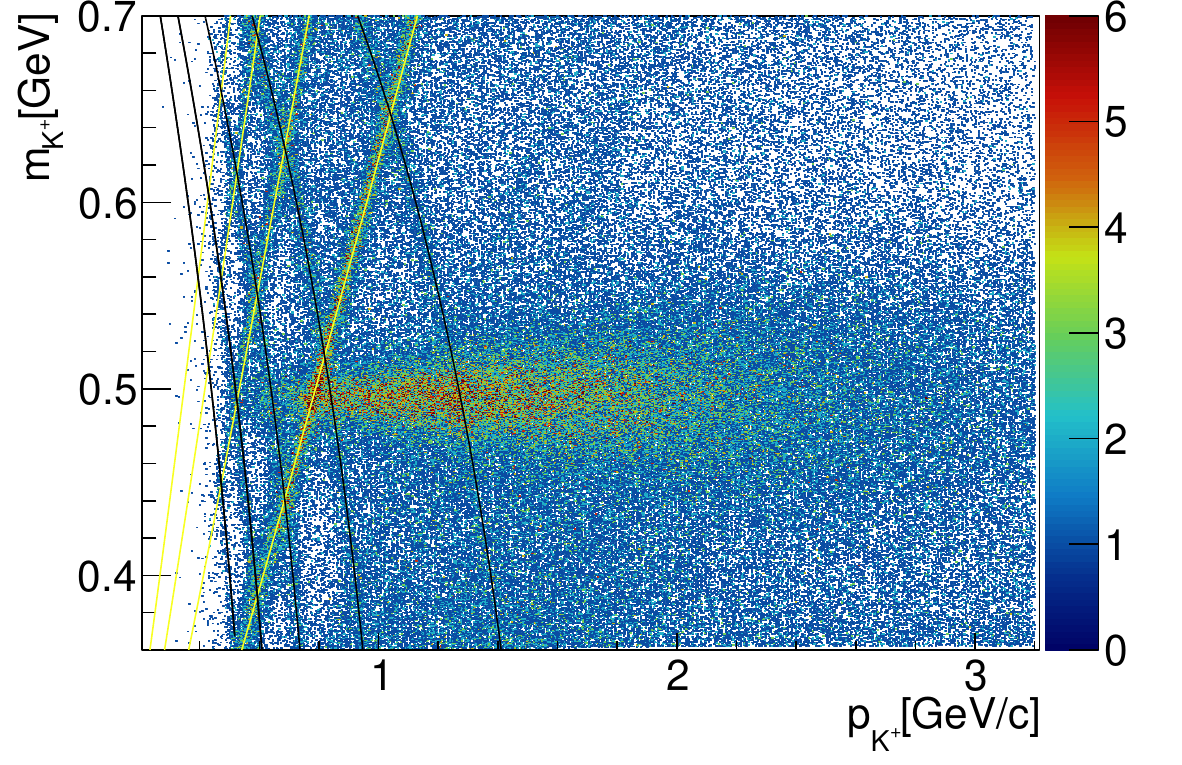}
\end{center}
\caption{Data taken in Fall 2018 with the \emph{inbending} torus magnet configuration. Kaon mass, $m_{K^{+}}$, plotted against the magnitude of kaon momentum, $p_{K^{+}}$, calculated from Eq.~\ref{kmass}. Out-of-time particle lines are highlighted by faint lines: yellow for pion and black for proton. }
\label{Crossings}
\end{figure}

If Eq.~\ref{kmass} is used to calculate the particle mass as a function of momentum, the resulting distribution shown in Fig.~\ref{Crossings} reveals a distinct kaon signal accompanied by smooth background contributions from pions (below kaon mass) and protons (above kaon mass). In addition, out-of-time particles appear as faint bands, highlighted by solid lines: black lines correspond to protons originating from subsequent bunches, while yellow lines represent pions from preceding bunches.

These additional components further illustrate the complexity of the background and the need for a more sophisticated background-suppression technique. Simple PID-$\chi^2$ cuts, regardless of how narrowly they are chosen, cannot eliminate these contributions. The kaon signal exhibits significant overlap with multiple background contributions, making reliable signal extraction challenging.

These misidentification structures and the shape of the kaon mass are momentum dependent in amplitude, shape and width. The production probability depends on the momentum of outgoing particles. The momentum resolution, $\frac{\Delta p}{p}$, in FD is nearly constant at a given angle. The timing resolution of the detector is a characteristic of the ToF counters. The higher the momentum of a given particle, the shorter its time-of-flight, so at a fixed time resolution, the effective ToF separation between species with the same momentum worsens and the background suppression gets extremely complex. 

\section{sWeights \label{sec::SWeights}}

One method that enables statistical separation of signal and background contributions is the sWeight (or sPlot) technique~\cite{Pivk:2005splot}. The sWeight method reconstructs the true distributions of signal and background components for a variable of interest (the control variable) by exploiting information from a separate discriminating variable in which signal and background can be modeled and distinguished.
A key assumption of the method is that the control variable is independent of the discriminating variable~\cite{Pivk:2005splot}. Under this assumption, event-by-event weights can be calculated from a fit to the discriminating variable and subsequently used to statistically unfold the signal and background contributions in the control variable.
Accurate Probability Distribution Functions (PDFs) of the signal and background in the discriminating variable are essential for the reliability of the method. Importantly, the technique is not limited to a simple signal–background scenario: it can accommodate an arbitrary number of species, provided that each component can be described by a distinct PDF and that the total model correctly represents their sum. For clarity, however, the present discussion will be restricted to the case of two species: signal and background. The total PDF, $f(x)$, can then be written in the form:

\begin{align}
    f(x) = N_S f_S(x) + N_B f_B(x)
\end{align}
where $N_S$ and $N_B$ are signal and background yield, $f_S$ and $f_B$ are the respective signal and background PDFs. The formula to calculate signal and background weights, $sP_S(y_e)$ and $sP_B(y_e)$, for the discriminating variable in event e, $y_e$, is given by:

\begin{align}
    sP_S(y_e) &= \frac{V_{S,B} f_{B}(y_e) + V_{S,S} f_{S}(y_e)}{N_B f_{B}(y_e) + N_S f_{S}(y_e)}
    \\
    sP_B(y_e) &= \frac{V_{B,S} f_{S}(y_e) + V_{B,B} f_{B}(y_e)}{ N_B f_{B}(y_e) + N_S f_{S}(y_e)},
\end{align}

where $ V_{nj}$ for $ n,j \in \{S,B\}$ is the covariance matrix element.

To know the total number of events in each species, a simultaneous maximum likelihood fit of background and signal is performed on the discriminating variable. From the fit, the parameters of the underlying PDFs and the total number of events in each species are extracted. The inverse covariance matrix elements are related to the second derivative of the likelihood function, $\mathcal{-L}$, with respect to the parameters at the central values from the fit,
\begin{align}
(V^{-1})_{nj}
=
\frac{\partial^2(-\mathcal{L})}{\partial N_n\,\partial N_j}
=
\sum_{e=1}^{N_{tot}}
\frac{f_n(y_e)\,f_j(y_e)}
{\left(\sum_{k=1}^{2} N_k\,f_k(y_e)\right)^2},
\end{align}
where $N_{tot}$ is the total number of events and the inner sum runs over the number of species. 
The observed distribution of $y$ for a case of two species is then given by:
\begin{align}
    f_{obs}(y) &= \frac{1}{N_S+ N_B} \sum^{N_{tot}}_e \delta(y- y_i) \nonumber
    \\
    & = \frac{N_S}{N_S+ N_B} f_S(y) + \frac{N_B}{N_S+ N_B} f_B(y) 
\end{align}
Each event is a mixture of signal and background. The covariance matrix corrects for correlations between yields of different species, ensuring on average, the weighted sum reproduces the true yield. By construction, the signal and background weights assigned to a given event sum to one.

\section{1D Fitting with sWeights \label{sec::1Dissues}}
Kaon mass can be utilized as the discriminating variable to perform sWeighting on a chosen distribution. However, because both the signal and background components of kaon mass depend strongly on momentum, the discriminating variable cannot be treated in a purely one-dimensional manner. One practical solution is to divide the dataset into momentum slices, which restores approximate one-dimensionality for each slice and allows kaon mass to be used effectively as the discriminating variable. The ranges of the slices are chosen to distribute the events as evenly as possible. Ideally, the slices should be narrow enough to reduce momentum-dependent effects, while still containing sufficient statistics to perform reliable fits. The kaon signal, $S(m)$, in each slice can be modeled as a Gaussian centered at kaon mass, $\mathrm{G_K}$, and the smooth in-time background, $B(m)$, as the sum of a Gaussian centered at the pion mass, $\mathrm{G_\pi}$, and a Gaussian centered at proton mass $\mathrm{G_{P}}$:
\begin{align} \label{PDFS}
    S(m) &= \alpha_1 \mathrm{G_K}(m, m_K, \sigma_1)
    \\
    B(m) &= \alpha_2 \mathrm{G_\pi}(m, m_\pi, \sigma_2) + \alpha_3 \mathrm{G_P}(m, m_P, \sigma_3) \nonumber,
\end{align}
\begin{figure*}[t]
\begin{center}
\includegraphics[width=0.44\textwidth,angle=0]{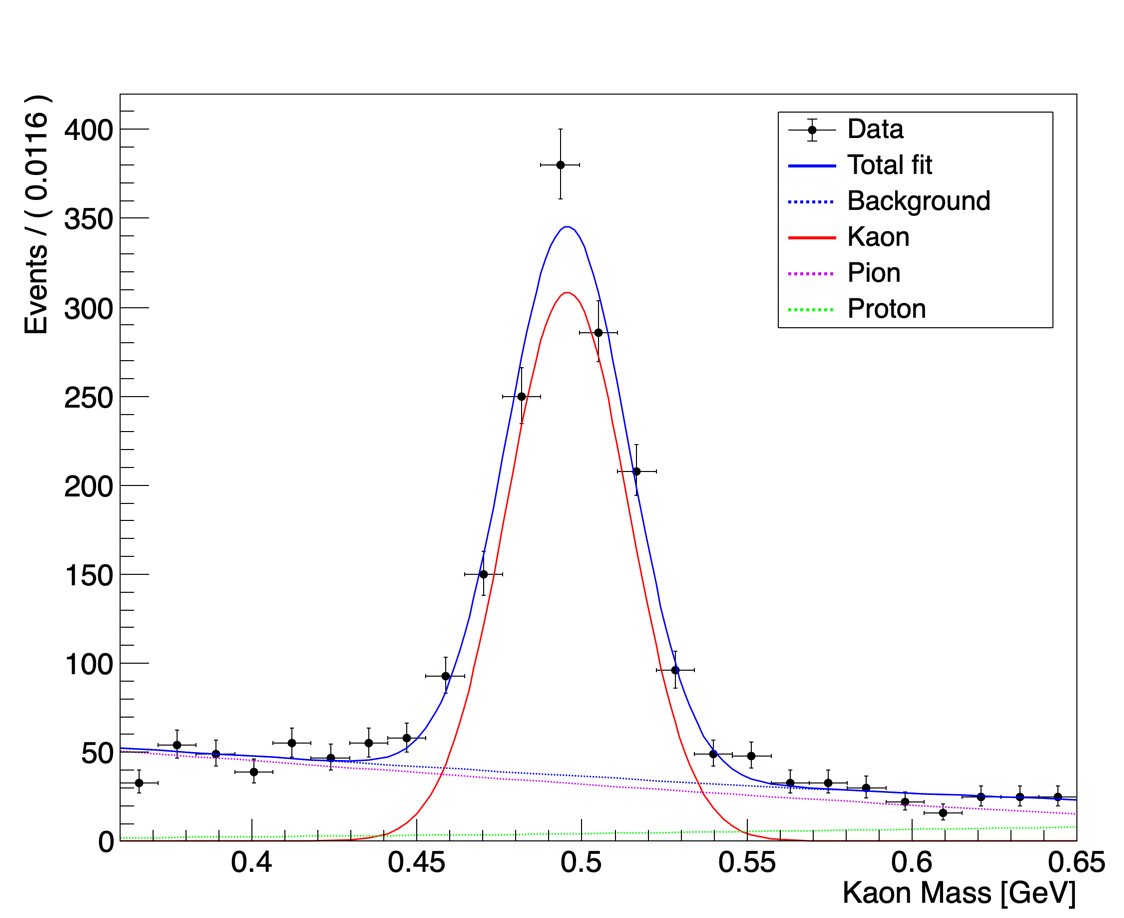}
\includegraphics[width=0.45\textwidth,angle=0]{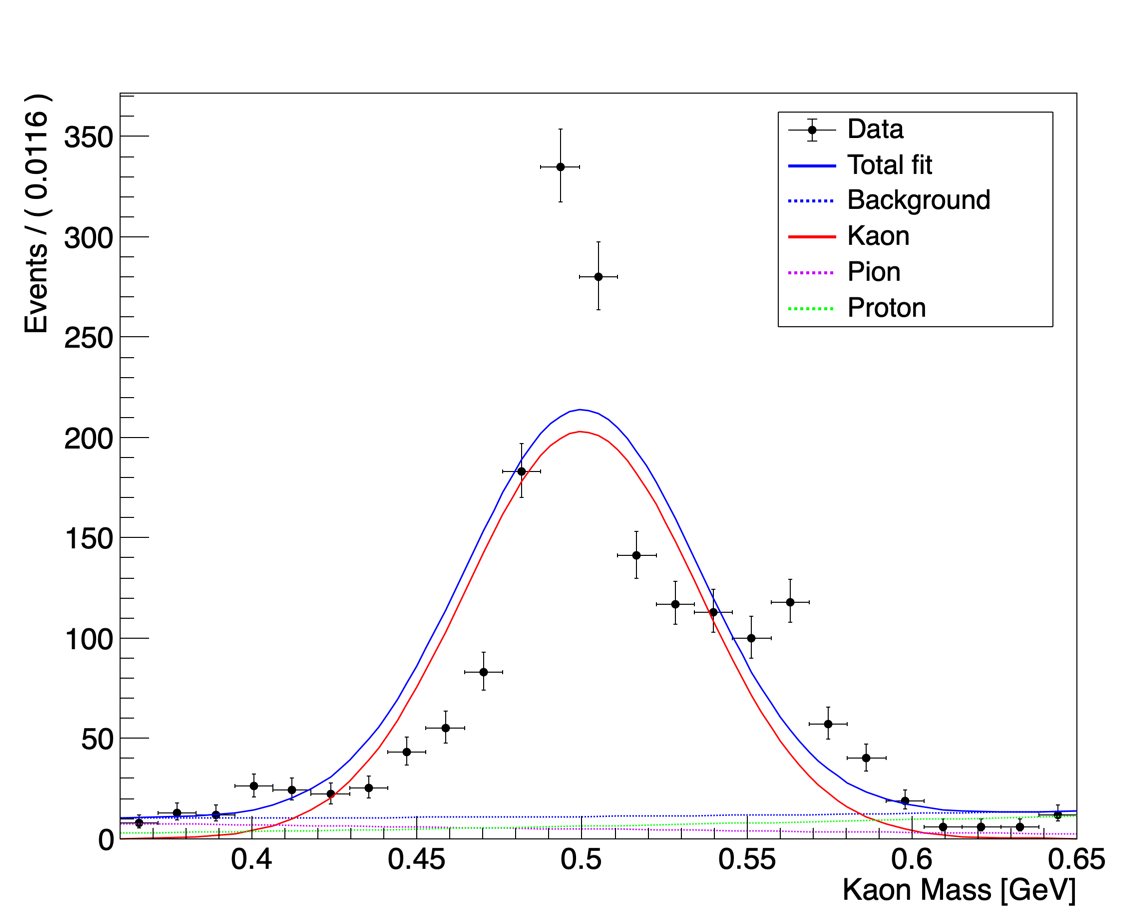}
\end{center}
\caption{Data taken in Spring 2019, with torus magnet in \emph{inbending} mode, kaon mass fits (detailed by Eq. \ref{PDFS}) for two different slices in momentum. Data is given in black points, total fit in solid blue, total background in dotted light blue and signal in red. The pion and proton components of the background are given in dotted magenta and green respectively. \underline{Left}: An example 1D fit where the kaon mass is fit well. Kaon momentum range $p_{K^+}\in[1.2, 1.3]~GeV/c$.  \underline{Right}: An example 1D fit slice where the contribution from the out-of-time particle contribution compromises the integrity of the fit, kaon momentum range $p_{K^+}\in[0.4, 0.8]~GeV/c$.}
\label{SplotSlicesCrossingsNotBeingFitted}
\end{figure*}
where $m$ is the calculated mass of the kaon in the event (running parameter in Gaussian distribution), and $m_{K,\pi,P}$ are the known masses of each respective particle specified in the subscript. $\sigma_{1,2,3}$ and $\alpha_{1,2,3}$ are parameters to be fitted, representing the width and scale of the Gaussians, respectively. 
This method was initially tested using data collected with CLAS12 during the Spring 2019 running period with the torus magnet operating in the \emph{inbending} configuration. To improve particle identification, additional selection criteria were applied using the CLAS12 Event Builder variable, $\chi^{2}_{\mathrm{PID}}$ ~\cite{Ziegler2020CLAS12}. For electrons and positrons, $\chi^{2}_{\mathrm{PID}}$ is defined as the deviation, expressed in units of the detector resolution, $\sigma$, between the measured electromagnetic calorimeter sampling fraction (the ratio of the total ECAL energy deposition to the particle momentum) and its expected value. For charged hadrons, several particle hypotheses are evaluated and the hypothesis that minimizes the difference between the reconstructed vertex time and the event start time is assigned. The corresponding value of $\chi^{2}_{\mathrm{PID}}$ is then defined as the timing residual, $\Delta t$, divided by the timing resolution, $\sigma$. Charged particle candidates were required to satisfy $|\chi^{2}_{\mathrm{PID}}| < 3$. An example of a successful fit on a momentum slice ($p_{K^+}\in[1.2, 1.3]~GeV/c$) is shown in Fig.~\ref{SplotSlicesCrossingsNotBeingFitted}~(Left). The imposed $\chi^{2}_{\mathrm{PID}}$ cuts on both kaons reduce out-of-time background in the tails of the kaon mass distribution but can not eliminate it in the central mass band. One can see that in Fig.~\ref{SplotSlicesCrossingsNotBeingFitted}~(Right) ($p_{K^+}\in[0.4, 0.8]~GeV/c$), the unaccounted for out-of-time bands can severely distort the fit.

\begin{figure}[H]
\begin{center}
\includegraphics[width=0.5\textwidth,angle=0]{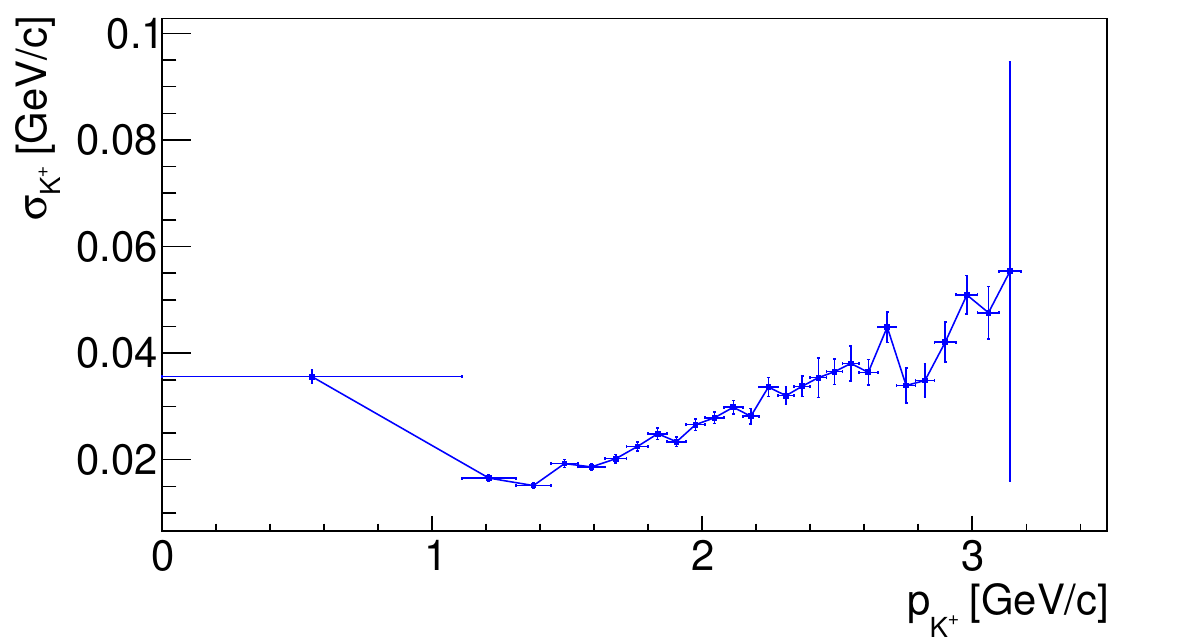}
\includegraphics[width=0.485\textwidth,angle=0]{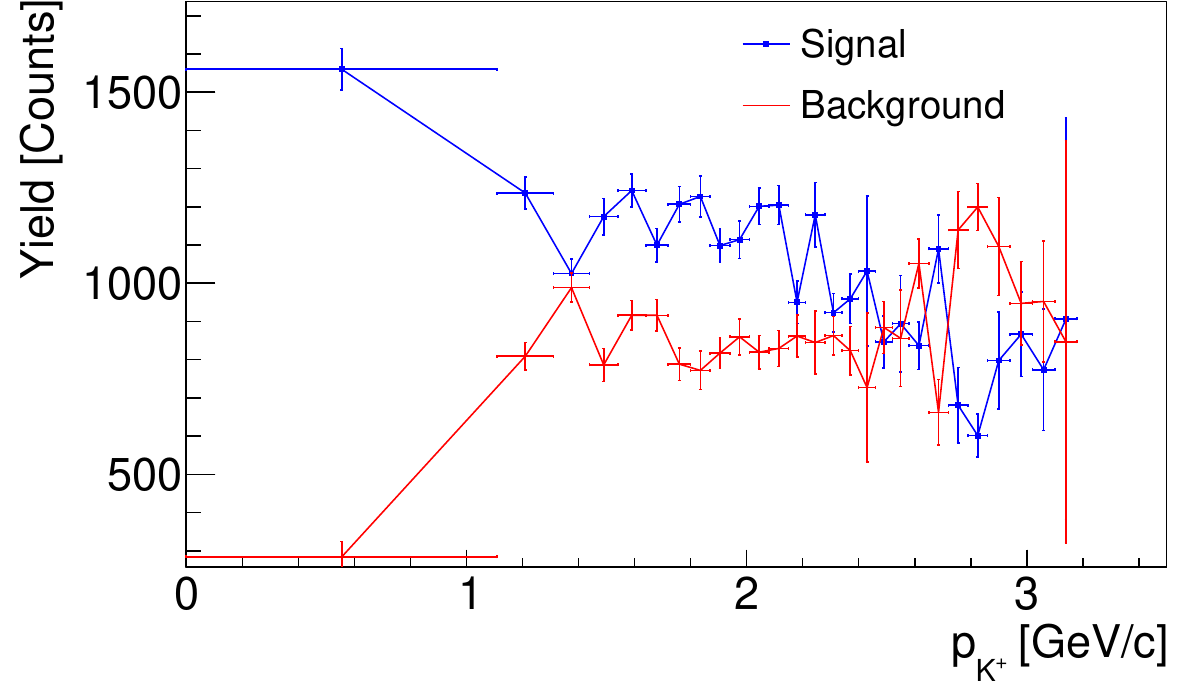}
\end{center}
\caption{Parameters from fits to data taken in Spring 2019, with the torus magnet in \emph{inbending} mode. \underline{Top}: Kaon width parameter for each slice over the full kaon momentum range. \underline{Bottom}: Signal (blue) and background (red) yield calculated using the sWeight method for each slice of momentum over the full kaon momentum range. One can see that the incorrect fits result in many false positive signal events.}
\label{KaonWidthBgSigHeight}
\end{figure}

Combining fits from different momentum slices, the momentum-dependent width of the kaon signal distribution can be plotted as shown in Fig.~\ref{KaonWidthBgSigHeight}~(Top). A rising trend is clearly visible despite some scatter. Limited statistics at small kaon momentum and additional background do not allow for a reasonable fit below 1~GeV/c.

While the signal width varies smoothly with kaon momentum, the signal yield fluctuates significantly from slice to slice, as shown in Fig.~\ref{KaonWidthBgSigHeight}, and appears anti-correlated with the background yield. The out-of-time background (i.e. crossings) only occurs within narrow momentum intervals rather than across the full range. This makes it extremely challenging, if not impossible, to include them consistently in a one-dimensional fit. As a result, the kaon mass distribution cannot be fully modeled in these slices, causing the fit to fail wherever prominent crossings are present, as shown in Fig.~\ref{SplotSlicesCrossingsNotBeingFitted} (Right).

To overcome these limitations and reduce systematic uncertainties associated with binning or slicing, the conventional one-dimensional sWeight technique was extended to a two-dimensional weighting approach.

\section{2D Fitting \label{sec::2D}}
A two-dimensional fit allows one to model the momentum dependence of both the signal and the background smoothly, without relying on slices. As shown in Fig.~\ref{Crossings}, all distributions exhibit complex shapes, which makes the choice of an appropriate fit function critical to the validity of the method. Additionally, the parametrization should avoid introducing artificial correlations between different momentum regions, which often occurs when using polynomial fits. To ensure well-defined ranges, most of the distributions are parametrized using a Gaussian sampling method~\footnote{This method has several other names in the literature: "Gaussian basis expansion on a uniform grid", "Uniform Gaussian basis representation", etc.}.

\subsection{Gaussian sampling} \label{chap:GausSample}
\begin{figure}[h!]
\begin{center}
\includegraphics[width=0.445\textwidth,angle=0]{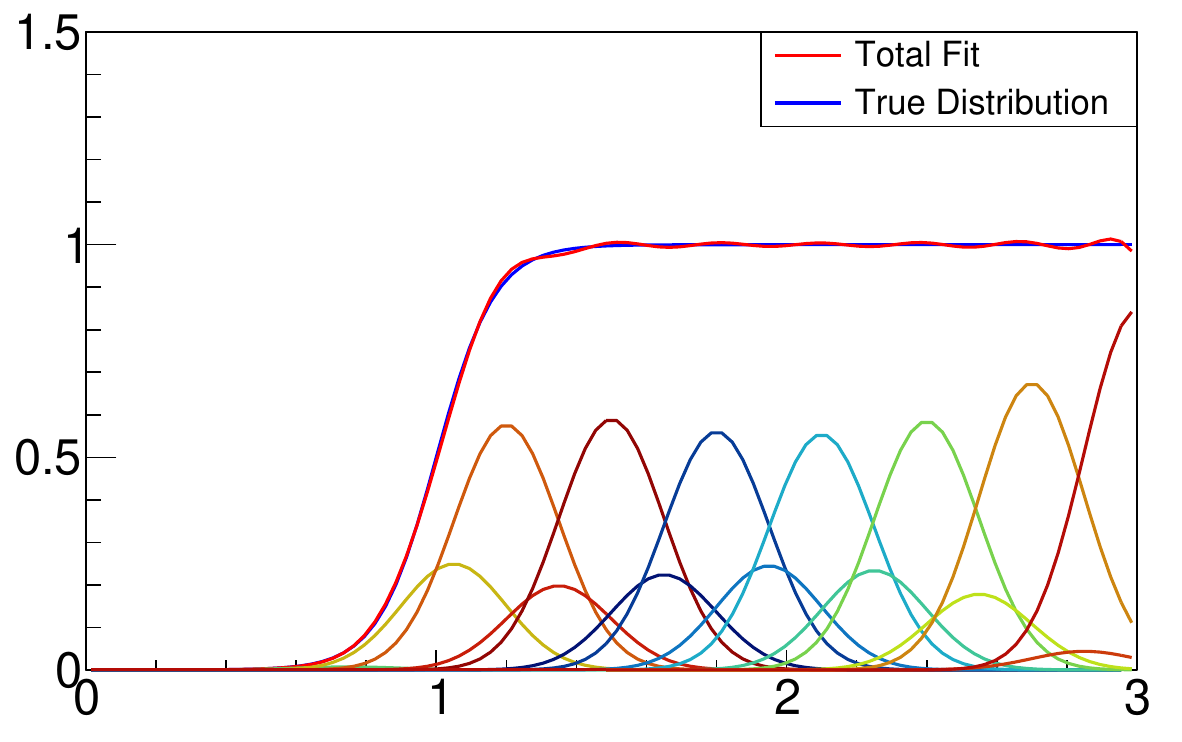}
\includegraphics[width=0.445\textwidth,angle=0]{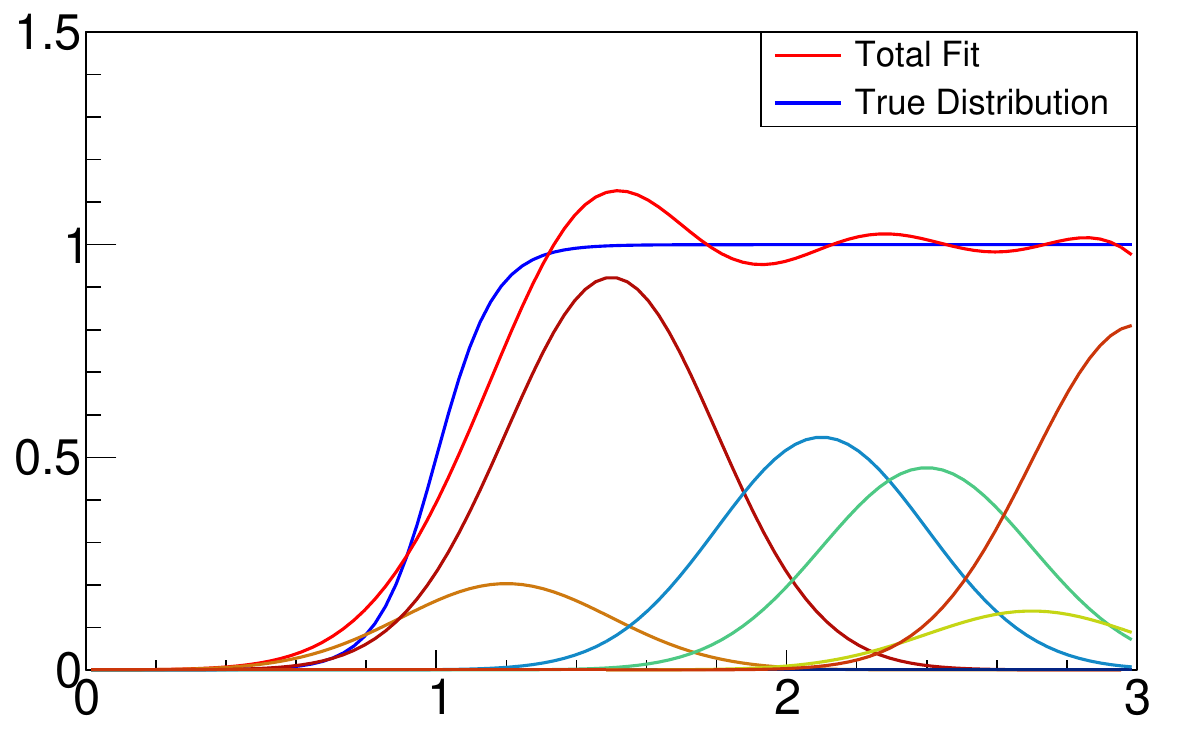}
\end{center}
\caption{Gaussian sampling (red) with two different Gaussian widths used to approximate a sigmoid function (blue). Colored Gaussians show individual contributions of the expansion. \underline{Top}: $\sigma = 0.15$~. \underline{Bottom}: $\sigma = 0.3$~.}
\label{GaussianSampling1}
\end{figure}

An analytic function can be represented as a sum of equidistant, overlapping Gaussians of the same width. The means of these Gaussians, $\mu_j$, are spaced by the width, $\sigma$, such that $\mu_j = \mu_0 + j*\sigma $ where $\mu_0$ is the mean of the first Gaussian, and $j$ is the total number of Gaussians. An approximation of a sigmoid function using this method is shown in Fig.~\ref{GaussianSampling1} as an example.

The width of these Gaussians controls the maximum resolvable gradient in the distribution, while the total number $j$ determines the range of interest. Distributions with sharp rises or falls require narrower Gaussians, as illustrated in Fig.~\ref{GaussianSampling1}. Conversely, slowly varying functions can be adequately approximated with fewer, wider Gaussians. Narrower Gaussians, however, require more parameters to accurately model the function, which directly increases the computational cost.

Gaussian sampling allows controlled variability and positive definition, while asymptotically going to  zero outside the range of interest - mathematical properties essential to define PDF's. 

\subsection{2D parametrization}
To benchmark the idea of multidimensional sWeight technique, RGA (Fall 2018 and Spring 2019) two-$K^+$ production data samples were utilized, where all final state particles (scattered electron and the two positively charged kaons) were detected in the FD, $e^- p\to e^- K^+K^+ X$. We deliberately choose to detect all particles in the same detector, FD, to minimize detector related reconstruction effects and maximize the method performance information. All $\chi^{2}_{\mathrm{PID}}$ requirements on kaons were removed to investigate the out-of-time background in detail and avoid introducing biases from particle identification requirements. From now on the faster kaon of the two will be referred to as kaon 1, and the slower one as kaon 2. The kaon signal, $S(m,p)$, can be modeled as a Gaussian centered at kaon mass, $\mathrm{G}(m, m_k, w(p))$, and the background, $B(m,p)$, as the sum of in-time pion and proton  background, $B_\pi(m,p), B_P(m,p)$ (see details in Section~\ref{chap::PiPr}) and out-of-time pion and proton background or crossings, $C_\pi(m, p, \Delta t), C_P(m, p, \Delta t)$ (details given in Section~\ref{chap::Cross}):
\begin{align} \label{Gaussians}
    S(m ,p) &= \alpha_1(p) \mathrm{G}(m, m_K, w(p))
    \\ \nonumber
    \\
    B(m, p) &= \alpha_2(p){B_\pi}(m,p) + \alpha_3(p){B_P}(m,p)
    \\
    & + \sum^{N_{\pi}}_{n=1} \alpha_{4_n}(p)C_\pi(m, p, \Delta t=4n)   \nonumber
    \\ \nonumber
    & +\sum^{N_{P}}_{n=1}  \alpha_{5_n}(p) C_P(m,p, \Delta t=4n) \nonumber
\end{align}
Where $\Delta t$ is the time difference between the assigned beam bunch and the correct one. The number of pion and proton crossings $N_{\pi}, N_{P}$ that appear in the 2D mass-momentum distribution depend on the momentum of the kaon. Since beam bunches are $4$ ns apart in this dataset, $\Delta t$ is in multiples of 4. All visible crossings should be accounted for. where width $w(p)$ is given by:
\begin{align} \label{GaussiansW}
   w(p) = \sqrt{T_0^2+T_1(p- \mu)^2}.
\end{align}
where $p$ is the three-momentum, $\mu=0.5~GeV$ is the lowest three-momentum value in the sample, $T_{0,1}$ are the parameters to be fitted~\footnote{a typical distribution to parameterize the momentum resolution of magnetic spectrometers}. Momentum-dependent intensity, $\alpha_i$, of both signal and background are parametrized with Gaussian sampling, Eq.~\ref{fitfunc}, and constrained by an unbinned log likelihood fit.
\begin{align} \label{fitfunc}
    \alpha_i &= \sum_{n=0}^n a_n \mathrm{G}(p, \mu +n\sigma, \sigma)
\end{align}
Here, $a_n$ represent the amplitude of each Gaussian, and $\sigma$ is the standard deviation. The value of $\sigma$ is chosen to match the rate at which the distribution height varies across momentum space, while the number of Gaussians $n$ is selected to cover the entire momentum range. For our experiment the width of $\sigma = 100$~MeV appeared to be sufficient, see Section~\ref{chap:GausSample} for more details. An initial estimate of the required Gaussian width can be obtained by examining the rate of change of the signal amplitude in momentum slices.

To reduce the number of free parameters in a global fit, the lowest momentum slice where the kaon peak is visible can be fitted first to constrain and fix the value of $T_0$.

\begin{figure}[h!]
\begin{center}
\includegraphics[width=0.49\textwidth,angle=0]{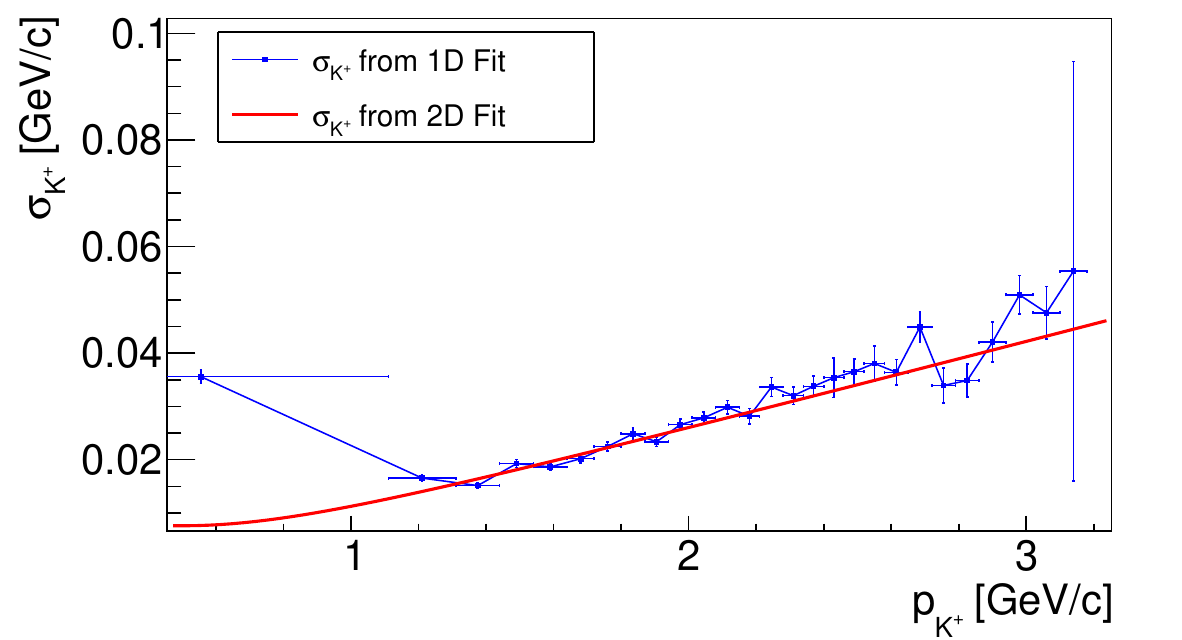}
\end{center}
\caption{Kaon signal width obtained from 1D slices fits (blue) and global 2D fits (red) with the width parametrization from Eq.~\ref{Gaussians}, extracted from fits to data taken in Spring 2019 with the torus magnet in the \emph{inbending} configuration.}
\label{KaonWidthParam1D2D}
\end{figure}

The kaon width increases regularly with momentum. Fig.~\ref{KaonWidthParam1D2D} shows a comparison between the widths extracted from the 1D slice fits and those obtained from the global 2D fit using the width parametrization of Eq.~\ref{GaussiansW}.
The result of a global 2D fit smoothly captures the rising dependence of the signal width, eliminating jumpy behavior present in the 1D fits.

\subsection{Crossings} \label{chap::Cross}
The momentum dependence of the mass for out-of-time particles, observed as crossings in Fig.~\ref{Crossings}, can be analytically determined from known beam bunch spacing. For a kaon with correctly calculated mass $m_K$, the velocity $\beta$ is related to its momentum $p$ by:

\begin{align}
    \beta^2=\left(\frac{d}{t}\right)^2 &=  \frac{p^2}{p^2 + m_K^2}=\frac{p^2}{E^2}
    \\
    \left(\frac{d}{p}\right)^2 &=  \frac{t^2}{p^2 + m_K^2}
    \\
    t^2 &= \left(\frac{d}{p}\right)^2 (p^2 + m_K^2) 
\end{align}
where $t$ is the time of flight and $d$ is the track length divided by speed of light. In the RGA, beam bunches arrive every 4 ns, so out-of-time particles have $t$ incorrectly assigned in multiples of 4 ns. The difference between a wrongly calculated particle mass $m_{\pi, P}$ and the true mass $m_K$ corresponds to $\Delta t$:

\begin{align}
    \left(\frac{d}{p}\right) \sqrt{(p^2 + m_K^2)} -  \left(\frac{d}{p}\right) \sqrt{(p^2 + m_{\pi, P}^2)} = \Delta t
\end{align}

This is in quadratic form. This equation can then be used to derive an expression for the calculated mass consistent with the particle’s true classification. 

Separate solutions for proton and pion can be obtained:
\begin{align}
    d_{p} = &-\Delta tp \frac{\sqrt{p^2 + m_P^2}}{m_{K}^2 - m_{P}^2}  \nonumber
    \\
    &+\sqrt{\frac{\Delta t^2p^2}{(m_K^2 - m_P^2)^2} (p^2 + m_P^2 ) + \frac{\Delta t^2p^2}{m_K^2 - m_P^2}}
\end{align}
\begin{align}
   d_{\pi} = &\Delta tp \frac{\sqrt{p^2 + m_{\pi}^2}}{m_K^2 - m_{\pi}^2}  \nonumber
   \\
   &+\sqrt{\frac{\Delta t^2p^2}{(m_K^2 - m_{\pi}^2)^2} (p^2 + m_{\pi}^2 ) + \frac{\Delta t^2p^2}{m_K^2 - m_{\pi}^2}}
\end{align}
The value of $d$ can be calculated analytically based on detector dimensions, but it is simpler to extract it from the data having $d$ as a fitting parameter. In this case $d$ was calculated to be ${\bf d=23.7}$ ns.
After implementing these relations in the fit, the linear parameter of the width component of the crossings, $T_1$, was determined to be in the order of $O(0)$~GeV indicating negligible momentum dependence of the crossing 
width, thus reducing the momentum-dependent width function to a simpler form:
\begin{align}
    C_{\pi}(m,p, \Delta t)= \nonumber
     \\ 
      \sum^{N_{\pi}}_{n=1} \alpha_{4_n} &\exp\left[-\frac{\left(m-\sqrt{\frac{\Delta t^2 p^2}{d^2} +m_\pi^2+\frac{2\Delta t p}{d}\sqrt{p^2+m_\pi^2}}\right)^2}{2\, \sigma_{C\pi}^2}\right] 
    \\
     C_{P}(m,p, \Delta t)= \nonumber
     \\ 
     \sum^{N_{\pi}}_{n=1} \alpha_{4_n}  & \exp\left[-\frac{\left(m-\sqrt{\frac{\Delta t^2 p^2}{d^2} +m_P^2 -\frac{2\Delta t p}{d}\sqrt{p^2+m_\pi^2}}\right)^2}{2\, \sigma_{CP}^2}\right]
\end{align}
The momentum dependence of the crossing widths, $\sigma_{C\pi, CP}$ is likely to be negligible due to appearance over a very narrow range of momentum.

\section{Sequential sWeights} \label{chap:seq}

When there is only one kaon per event, the sWeighting procedure is straightforward: 2D signal and background PDFs are created as described above, and the weights obtained from the unbinned log-likelihood fit are used to quantify the signal and background yields. The situation becomes more complex when multiple kaons are present in a single event. In the analysis of cascade production, for example, two positively charged kaons appear in the final state. To quantify signal and background in this case, a single weight per event is required to probabilistically attribute the event to either signal or background. There are several ways to extend sWeighting from the single-kaon case to multiple kaons. In this chapter, we describe various attempts to generalize 2D sWeighting to events with multiple kaons, ultimately presenting a working solution.

\subsection{4D weighting}
In previous chapters, we demonstrated how sWeighting can be extended from 1D (Eq.~\ref{Gaussians}) to 2D (Eq.~\ref{Gaussians}) cases. In general, the probability of finding an event can always be defined to sum to 1, consisting of contributions from signal and background. In the case of $2K^+$ production, there are four possibly correlated discriminating variables that define the quality of an event: $p_{k_1}$, $m_{k_1}$, $p_{k_2}$, and $m_{k_2}$. Consequently, the signal and background probability distribution functions, $S_{4D}, B_{4D}$, can be expressed as functions of these four variables, as shown in Eq.~\ref{Eq:Gauss_4D}.
\begin{align} \label{Eq:Gauss_4D}
    f_{4D}&(p_{[k_1,k_2]},m_{[k_1,k_2]}) = \nonumber
    \\ 
    &N^{frac}_S f_{S}(p_{[k_1,k_2]},m_{[k_1,k_2]}) +N^{frac}_B f_{B}(p_{[k_1,k_2]},m_{[k_1,k_2]}) 
\end{align}
Where:
\begin{align}
    N^{frac}_S = \frac{N_S}{N_S+ N_B}, N^{frac}_B = \frac{N_B}{N_S+ N_B}
\end{align}
The signal and background PDFs are individually normalized to unity, such that the total PDF is also normalized to unity.
While mathematically straightforward, several challenges arise when applying this approach to the current problem. Parametrizing a 4D surface with an arbitrary function is extremely difficult if the shape is not known {\it a priori}. The number of parameters required for such a function becomes prohibitively large, making the fitting procedure unfeasible, and the amount of data needed to constrain the fit far exceeds what is available from a single beamtime. Because a full 4D fit is impractical, simplifications are necessary to make the method viable for typical analysis.

\subsection{General 2D weighting and issues}

\begin{figure}[h!]
\begin{center}
\includegraphics[width=0.4\textwidth,angle=0]{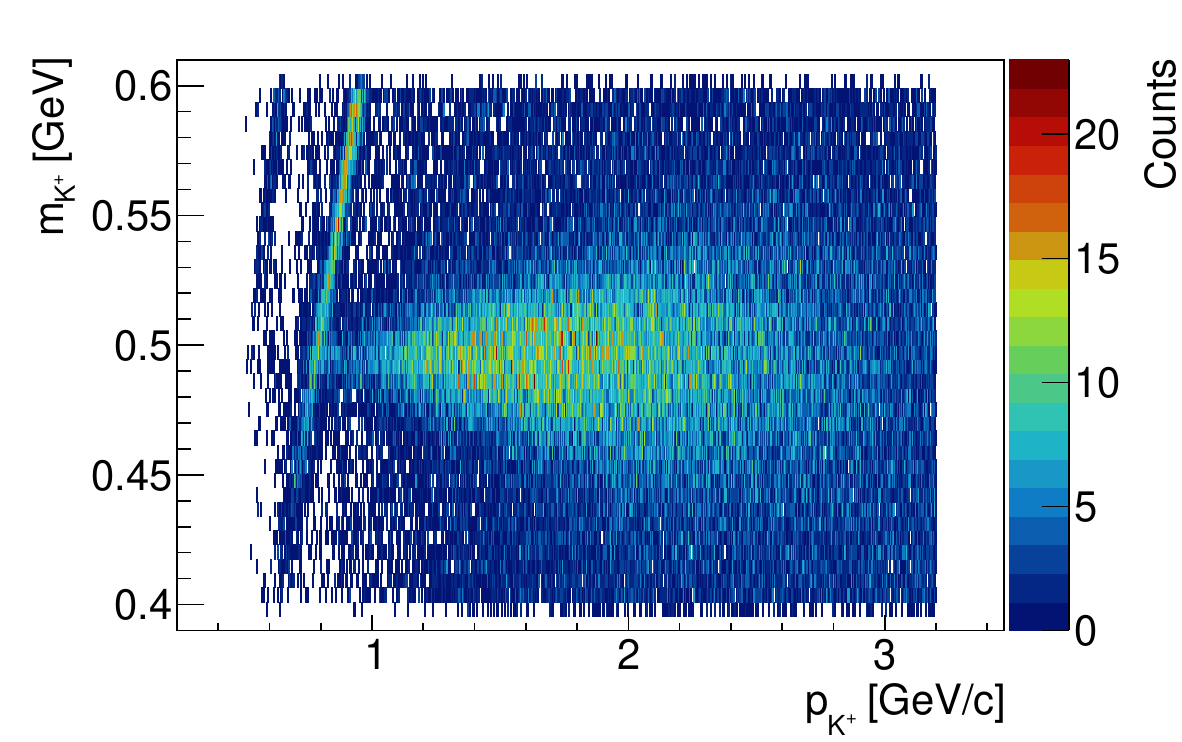}
\includegraphics[width=0.4\textwidth,angle=0]{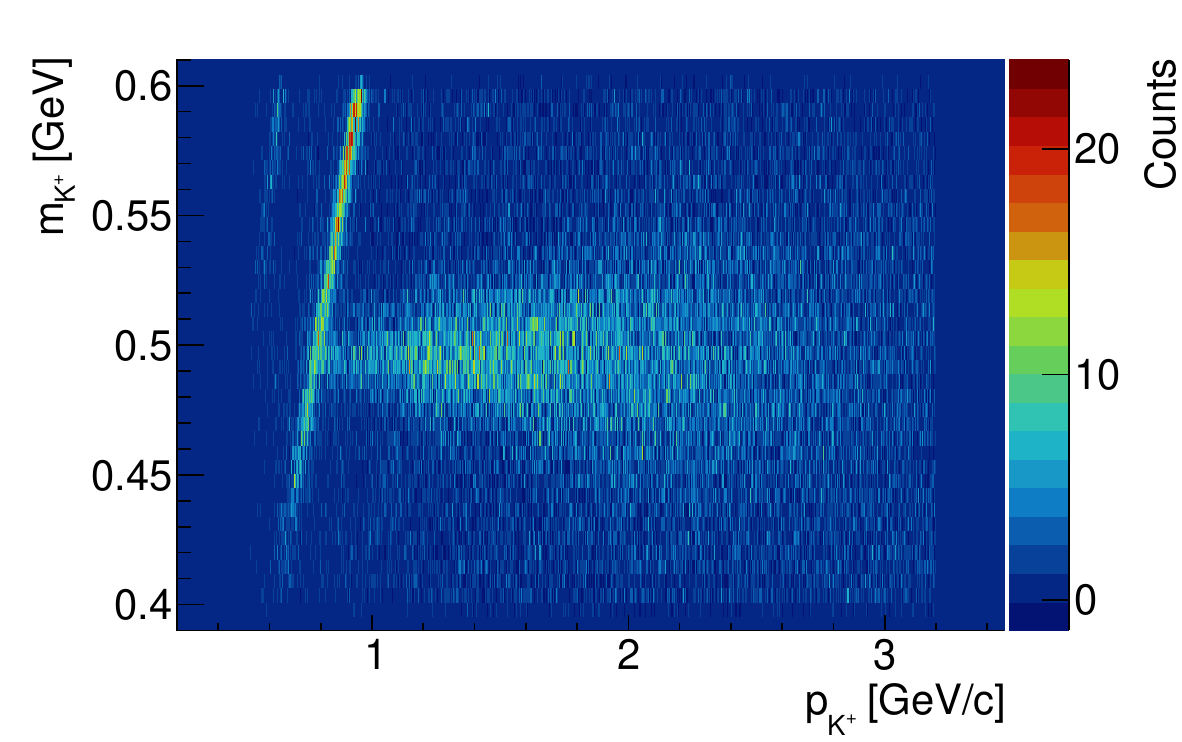}
\end{center}
\caption{Data taken in Spring 2019 with the torus magnet in the \emph{inbending} configuration. \underline{Top}: Unweighted kaon 1 mass vs momentum distribution. \underline{Bottom}: Kaon 1 mass vs momentum distribution, weighted with  kaon 2 background weights, enhancing the out-of-time crossings in kaon 1.}
\label{EnhanceCross}
\end{figure}

A very natural extension of 4D sWeighting would be a factorization approach. In sWeights the number of species do not have to be limited to two. One can extend the sWeight equation to $N_s$ species:
\begin{align}
    sP_n(y_e) = \frac{\sum^{N_s}_{j = 1} V_{n,j} f_{j}(y_e)}{\sum^{N_s}_{k = 1} N_k f_{k}(y_e)}
\end{align}
Where $N_k$ is the average number of events expected for the $k^{th}$ species, $f_k$ is the PDF of the discriminating variable for the $k^{th}$ species, $y_e$ is the discriminating variable for event e, and $ V_{n,j}$ is the covariance matrix element. The inverse covariance matrix is related to the weights:
\begin{align}
(V^{-1})_{nj}
=
\frac{\partial^2(-\mathcal{L})}{\partial N_n\,\partial N_j}
=
\sum_{e=1}^{N_{tot}}
\frac{f_n(y_e)\,f_j(y_e)}
{\left(\sum_{k=1}^{N_s} N_k\,f_k(y_e)\right)^2},
\end{align}
If for the moment we assume that the correlations between kaon 1 and kaon 2 are negligible we can decompose $f_{4D}$:
\begin{align} \label{Eq:Gauss_2D_fac}
     &f_{4D}(p_{[k_1,k_2]},m_{[k_1,k_2]}) = \nonumber
     \\ \nonumber
     &\left(N^{frac}_S f_S(p_{k_1},m_{k_1})+ N^{frac}_B f_B(p_{k_1},m_{k_1}) \right)\cdot \nonumber
     \\ 
     &\left(N^{frac}_S f_S(p_{k_2},m_{k_2})+ N^{frac}_B f_B(p_{k_2},m_{k_2}) \right) \nonumber
     \\ \nonumber
     \\ \nonumber
    &=N^{frac}_{SS}f_S(p_{k_1},m_{k_1})f_S(p_{k_2},m_{k_2}) 
    \\ \nonumber
    &+N^{frac}_{SB} [f_B(p_{k_1},m_{k_1})f_S(p_{k_2},m_{k_2}) + f_S(p_{k_1},m_{k_1})f_B(p_{k_2},m_{k_2})] \nonumber
    \\
    & + N^{frac}_{BB}f_B(p_{k_1},m_{k_1})f_B(p_{k_2},m_{k_2})
\end{align}
Where:
\begin{align}
    N^{frac}_{SS} = \frac{N^2_S}{N_S+ N_B},N^{frac}_{SB}= \frac{N_SN_B}{N_S+ N_B}, N^{frac}_{BB}= \frac{N^2_B}{N_S+ N_B}
\end{align}
This formulation results in four possible contributions: both kaon candidates belong to the signal, both belong to the background, or either the first or second kaon event comes from the background. In principle, it would be convenient to define a single universal kaon signal/background function and apply it with two sets of input parameters (mass and momentum) for each kaon. Unfortunately, this approach fails in practice if kaon 1 and kaon 2 distributions are correlated. The problem is easier to illustrate using the background: if one of the ``kaons'' is actually an out-of-time particle from a crossing (see Fig.~\ref{Crossings}), the probability that the other particle also originates from a crossing increases significantly, because the background of the two kaons are correlated with each other. In such cases, the assumption of independence between the backgrounds in the factorized approach leads to biased results. This effect can be seen in Fig. ~\ref{EnhanceCross}, where allocation of kaon 2 in an out-of-time band leads to substantial change of kaon 1 distribution (Fig.~\ref{EnhanceCross}, Bottom) making the crossing line much more pronounced than in an unweighted case (Fig.~\ref{EnhanceCross}, Top).
\subsection{Sequential 2D weighting with separate fitting}
To address the limitations of the factorization scheme discussed in the previous section, a sequential weighting procedure is implemented, in which the first and second kaons are parametrized separately for signal and background, as shown in Eq.~\ref{Eq:Gauss_2D_sec}.
\begin{align} \label{Eq:Gauss_2D_sec}
    f_{4D}(p_{[k_1,k_2]},&m_{[k_1,k_2]})=f_{1}(p_{k_1},m_{k_1})
f_{2|1S}(p_{k_2},m_{k_2}) \nonumber
    \\
    &= \left(N^{frac}_S f_S(p_{k_1},m_{k_1})+ N^{frac}_B f_B(p_{k_1},m_{k_1}) \right) \cdot \nonumber
    \\
    &\left(\frac{N_{2S|1S}}{N_{1S}}f_{SS}(p_{k_2},m_{k_2})
    +\frac{N_{2B|1S}}{N_{1S}}f_{SB}(p_{k_2},m_{k_2}) \right) 
\end{align}
where $f_{2|1S}(p_{k_2},m_{k_2})$ is the kaon 2 PDF for events that are weighted by kaon 1 signal weights. Here, $N_{2S|1S}$ and $N_{2B|1S}$ denote the effective signal and background yields for kaon 2 in the sample selected by the kaon 1 signal weights, respectively, while $N_{1S}$ is the signal yield extracted from the kaon-1 fit. They satisfy:
\begin{align}
N_{2S|1S}+N_{2B|1S}=N_{1S} 
\end{align}
 This approach can be further generalized to more than two step weighting Eq.~\ref{Eq:Gauss_ND}. 
\begin{align} \label{Eq:Gauss_ND}
    f_{4D}(p_{[k_1,k_2]},&m_{[k_1,k_2]})=f_{1}(p_{k_1},m_{k_1})
f_{2|1S}(p_{k_2},m_{k_2})f_{3|1S|2S}(x)
\end{align}
by adding a further weighting step on a separate discriminatory variable $x$ weighted by signal weights from the previous steps, described by PDF denoted by $f_{3|1S|2S}(x)$. Here, the signal and the background on each consecutive step can be a function of very different variables, and the fitting procedure to extract PDFs should be performed with the weights from all previous steps applied. As a result, the procedure becomes order-dependent in practice: applying the weighting sequence $(K_1 \rightarrow K_2)$ generally does not yield results identical to those obtained from $(K_2 \rightarrow K_1)$. While the differences are often small, the intermediate mass--momentum distributions used in the fits can differ substantially depending on which kaon has been weighted first.

\begin{figure*}[t]
\centering
\includegraphics[width=0.4\textwidth,angle=0]{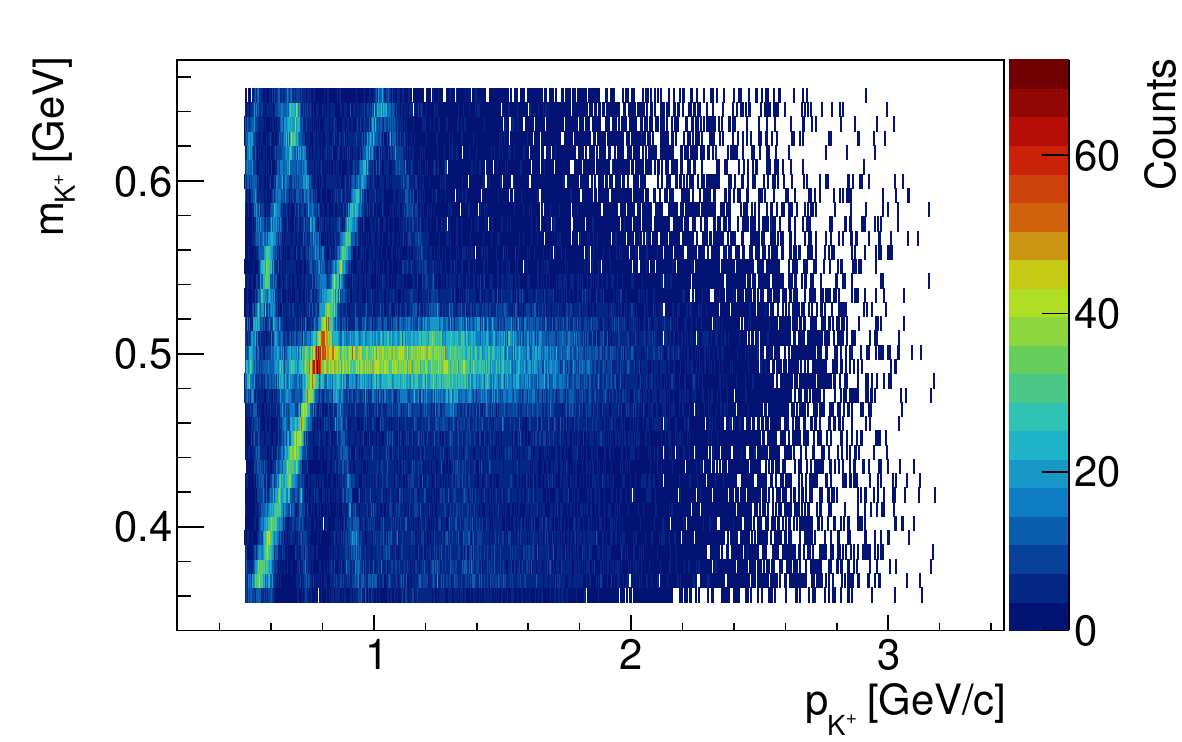}
\includegraphics[width=0.4\textwidth,angle=0]{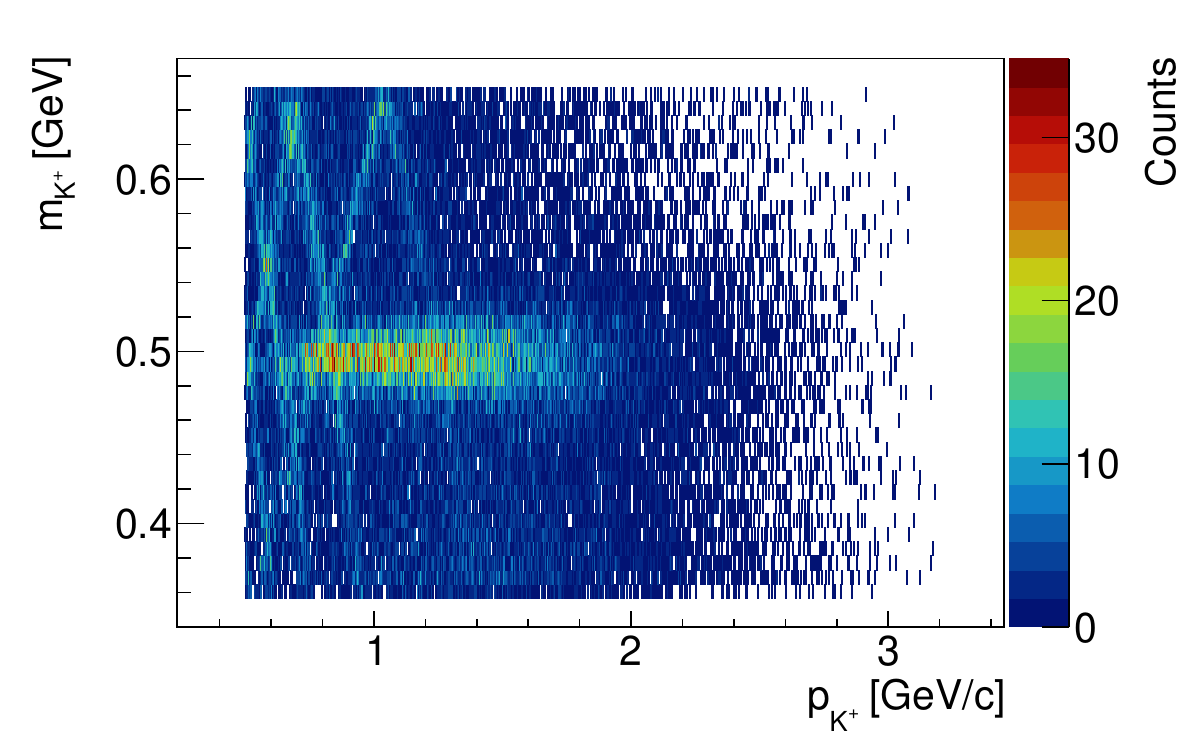}
\includegraphics[width=0.4\textwidth,angle=0]{h2DK1NoWeights.pdf}
\includegraphics[width=0.4\textwidth,angle=0]{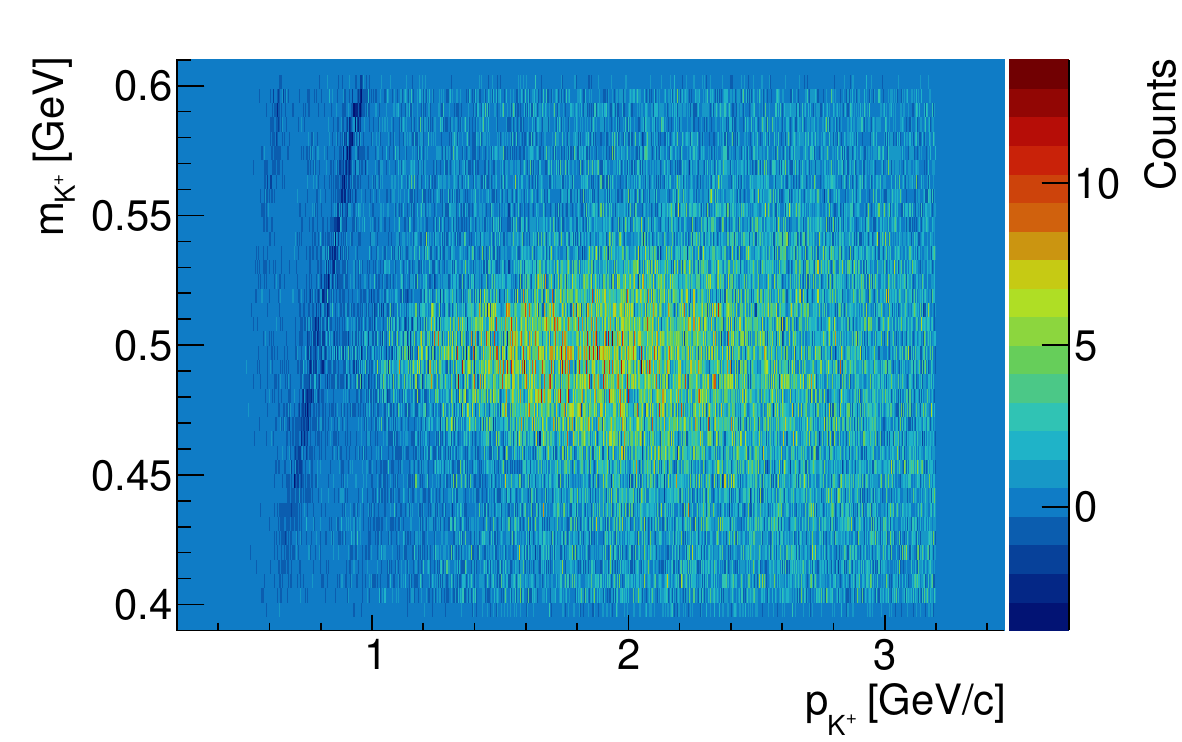}
\caption{Data taken in Spring 2019 with the torus magnet in the \emph{inbending} configuration. \underline{Top left}: kaon 2 unweighted. \underline{Top right}: kaon 2 weighted with kaon 1 signal weights. \underline{Bottom left}: kaon 1 unweighted. \underline{Bottom right}: kaon 1 weighted with kaon 2 signal weights. In the regions where the signal and background PDFs overlap, the crossings in this case, the weights become negative for many events.  Although negative weights are expected in the sWeight formalism, excessive overlap leads to large weight fluctuations and less reliable yield extraction.}
\label{K2WeightedAndUnweighted}
\end{figure*}

In the ideal limit of perfectly modelled probability density functions and negligible correlations, the final weighted distributions should be independent of the weighting order. In practice, however, finite statistics, imperfect parametrizations, and residual correlations can lead to different levels of fit stability and background suppression. Additional complications arise from correlations between the two kaons. For example, if the second kaon, which has many out-of-time crossings, is weighted first and the resulting weights are applied to the first kaon (see Fig.~\ref{K2WeightedAndUnweighted} Bottom row), the pion-crossing cross-correlations can lead to negatively weighted bins in the distribution. Conversely, if the first kaon (the faster one with fewer crossing lines) is weighted first and this weight is applied to the second kaon (see Fig.~\ref{K2WeightedAndUnweighted}, Top row), the signal is strongly enhanced over the background, and all events remain positively defined.

\subsubsection{Sequential 2D weighting - How-tos}

The resulting sequential procedure will appear as follows. As discussed earlier, kaon 2 provides greater signal--background discrimination than kaon 1. Consequently, the analysis proceeds from the weaker discriminator to the stronger one, progressively enhancing the signal purity at each stage. The weights extracted from the unweighted kaon 1 dataset, together with the kaon 2 dataset weighted using the kaon 1 signal weights, are used both in the construction of the weighted distributions and in the likelihood function for the extraction of the second-stage signal and background yields. The final signal distribution is constructed by applying the combined event weight to the control variable, $(MM(K^+K^+e^-))$. Formally, the final event weight is given by the product of the first-stage signal weight and the second-stage signal weight, where the latter is derived from a dataset already weighted by the former. Now, this is a delicate procedure. It is important to exercise caution when interpreting weights: they are not probabilities, and multiplying independent weights from separate fits does not produce a physically meaningful estimator. This is why a sequential approach is needed. In this framework, the second weighting stage is performed as a new signal extraction on the distribution obtained after the first weighting stage, preserving the statistical interpretation of the weighted sample and the final results can safely be used for yield extraction. This is not the same as multiplying two weights that have been obtained from independent fits. The validity of this approach depends on the degree of correlation between the variables used in successive weighting stages. Residual correlations may introduce biases in the extracted signal distributions, motivating the dedicated correlation studies presented in Section~\ref{sec::corr}.

\section{Data-driven approach in parametrization of backgrounds: pion and proton in-time background extraction based on $p\pi^+\pi^-$ exclusive events} \label{chap::PiPr}

Since the performance of the method is highly sensitive to the accuracy of the background parametrization, it is preferable to determine the in-time background shapes directly from the experimental data rather than relying on assumed functional forms. In our benchmark example, the shape of the in-time proton and pion backgrounds can be extracted by utilizing an independent reaction $ep \rightarrow e' p \pi^+ \pi^-$.

\begin{figure}[h!]
\begin{center}
\includegraphics[width=0.5\textwidth,angle=0]{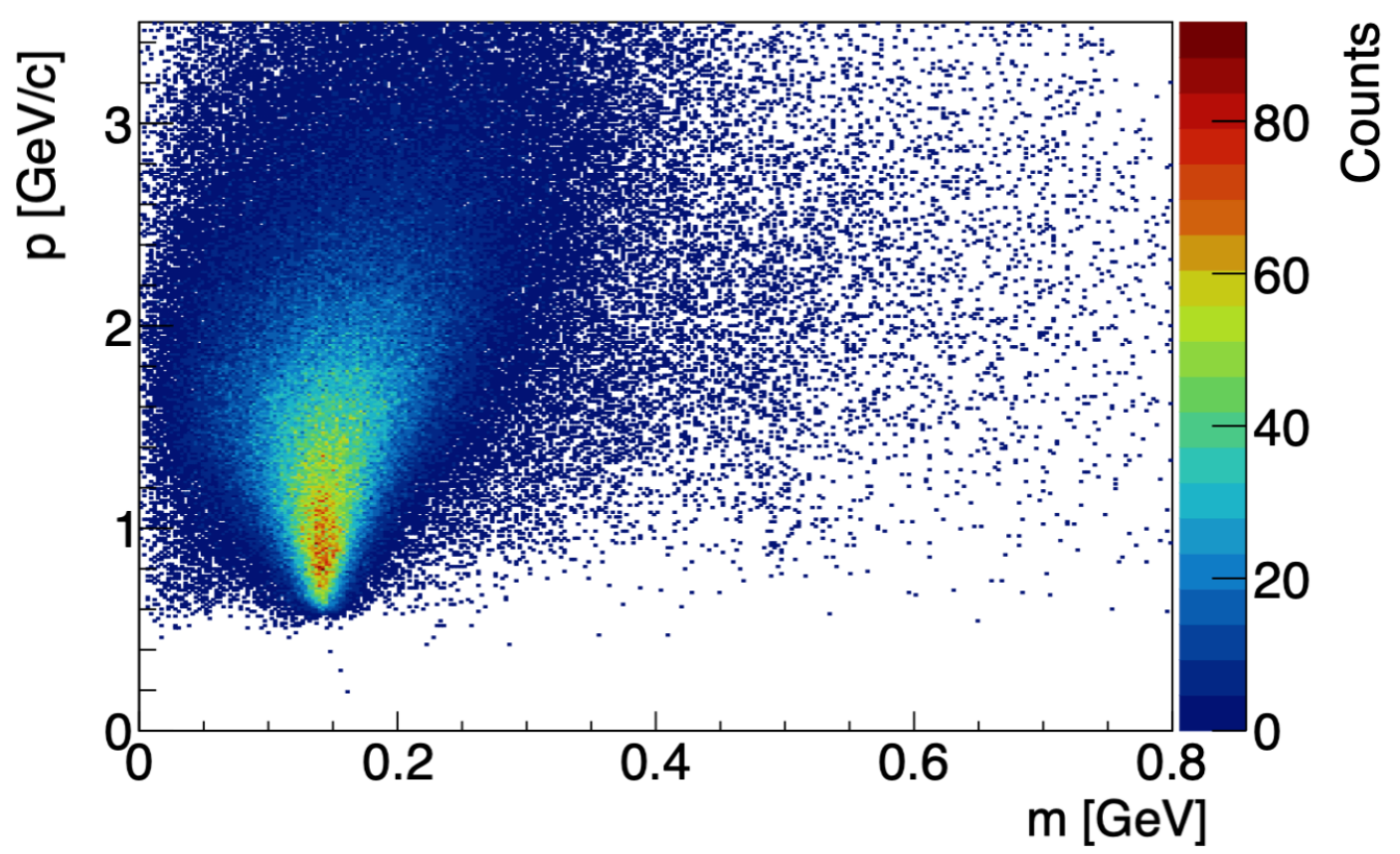}
\end{center}
\caption{Data taken in Fall 2018 with the torus magnet in the \emph{outbending} configuration. Momentum versus mass of $X^+$ in the reaction $ep \rightarrow e' p X^+ \pi^-$, where $X^+$ is a positively charged particle.}
\label{PionTail2D}
\end{figure}

\begin{figure}[h!]
\centering
\includegraphics[width=0.49\textwidth,angle=0]{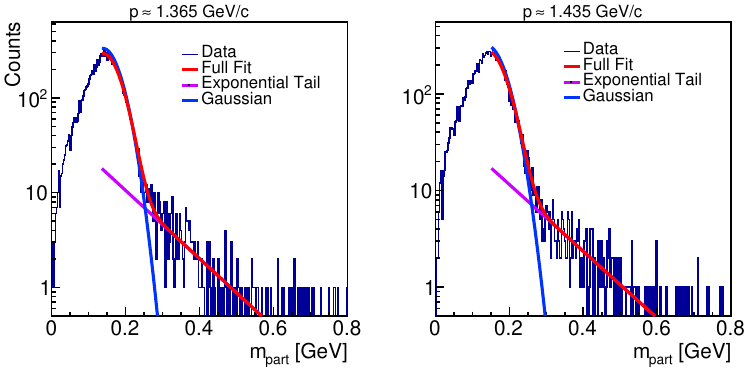}
\caption{Data taken in Fall 2018 with the torus magnet in the \emph{outbending} configuration, given as the blue histogram. Mass of $X^+$ in the reaction $ep \rightarrow e' p X^+ \pi^-$, where $X^+$ is a positively charged particle. Fits to extract pion tails in momentum slices, where the full fit is given in red, exponential and gaussian components of the fit are given in purple and blue respectively.  }
\label{PionTail1DFits}
\end{figure}

Events for this reaction were selected from RGA data taken in Fall 2018 with the torus magnet in \emph{outbending} mode comprising of exactly one proton, one electron, one $\pi^-$, and one additional positively charged particle in the final state to define the pion tail. Alternatively, events containing one electron, one $\pi^-$, one $\pi^+$ and one additional positively charged particle in the final state were used to define the proton tail. To identify this positively charged particle as a $\pi^+$, a cut to the opening angle between the missing momentum of the $e' p \pi^-$ system and the positively charged particle was applied. If the positively charged particle was a $\pi^+$, this opening angle is expected to be small. Additional constraints were imposed to the missing mass of the electron--proton system, motivated by the expectation that the $\pi^+\pi^-$ pair originates from a $\rho$ meson. Events are therefore loosely selected within the $\rho$ peak by requiring\[MM(e' p) < 1~\text{GeV}.\] After applying these selection criteria, a two-dimensional distribution of the positively charged particle mass versus its momentum was constructed, Fig.~\ref{PionTail2D}. The mass distribution was then examined in slices of momentum, and each distribution is fitted with a Gaussian centered at the pion mass together with an exponential tail component to model the background. The initial fitting function, $B_\pi(m,p)$ had the form:
\begin{align}
  B_\pi(m,p) =\alpha_2(p)\Big(e^{-C(p)[m-m_{\pi}]} + R(p)\cdot G(m, m_\pi, \sigma_{\pi}(p))\Big),
\end{align}
where $C(p)$ describes the shape of the exponential tail, and $m_{\pi}$ is the pion mass. The ratio of the Gaussian height to the exponential tail height is defined as $R(p)$. The width of the Gaussian is given by $\sigma_{\pi}(p)$.
From the fit slices shown in Fig.~\ref{PionTail1DFits}, it is clear that modeling the Gaussian signal component of the distribution is essential, as it extends into the kaon mass region.

To obtain the final form of the pion background distribution, an iterative fitting procedure was used. First, the ratio of the Gaussian height to the exponential height was determined by examining the fit parameters across the momentum slices. The component describing the functional momentum dependence of $R(p)$ was extracted and then fixed in the fitting function.

Next, the data were re-fit using the updated fitting function with the form of $R(p)$ fixed, allowing the functional dependence of $\sigma_{\pi}(p)$ to be determined. The extracted form of $\sigma_{\pi}(p)$ was then fixed in the fitting function, and the data were re-fit once more to determine the functional dependence of $C(p)$.

After this procedure, the resulting fitting function contains only one free parameter, $\alpha_2$. With the pion background now modeled, we can repeat the same steps to parameterize the shape of the proton background.

Events were selected from the RGA Fall 2018 \emph{outbending} dataset that contain exactly one electron, one $\pi^+$, one $\pi^-$ and one positively charged particle, and the  procedure described above in the pion tail extraction section is applied to extract our proton background. In the proton case, the signal peak is very narrow and very far away from the kaon mass range, so it is fair for the fitting function to consist of an exponential alone, ignoring the Gaussian signal: 
\begin{align}
  B_p(m,p) = \alpha_3 e^{C(p)(m-m_P)},
\end{align}
where $C(p)$ describes the shape of the exponential tail, and $m_{P}$ is the proton mass. The same iterative procedure was employed. 

\begin{figure}[H]
\begin{center}
\includegraphics[width=0.45\textwidth,angle=0]{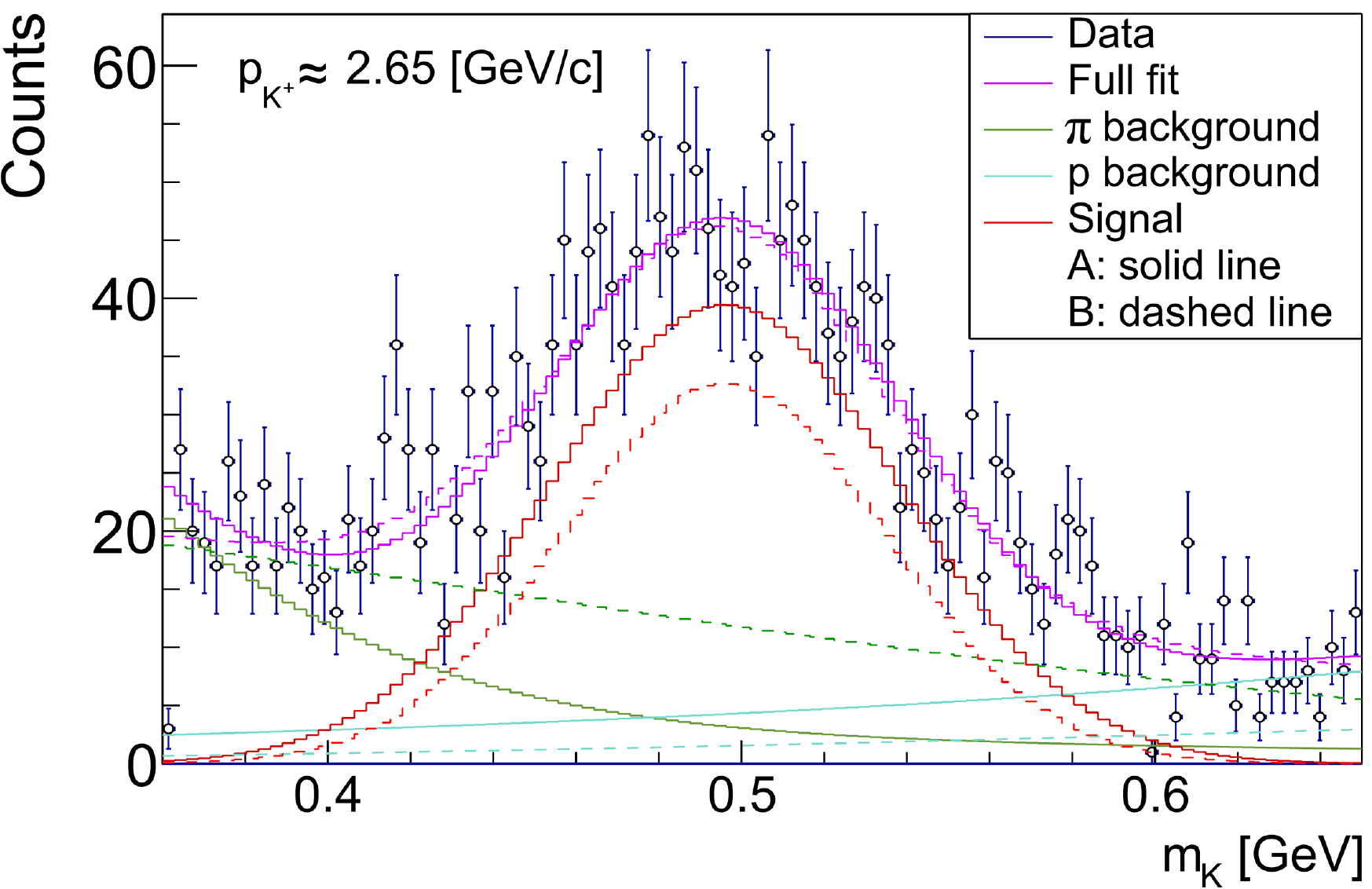}
\end{center}
\caption{Data taken in Fall 2018 with the torus magnet in the \emph{outbending} configuration given in blue data points. Comparison of overall fit and its components, before extraction of tails (B) in dashed lines and after (A) in solid lines. The full fit is given in magenta and the signal component is given in red curves. The pion and proton backgrounds are given in green and light turquoise respectively. }
\label{CompFitsPiTail}
\end{figure}

There is a significant difference between the usual fit, which assumes arbitrary Gaussian tails and the fit that is informed of the pion and proton background shape, especially in the high momentum regions, where the large width of the kaon signal does not allow to identify the exact shape of the background underneath,  Fig.~\ref{CompFitsPiTail}.

Fig.~\ref{CompFitsPiTail} demonstrates the outcome of such data-driven background shape extraction,  exhibiting an enormous difference in fits when the shape of the background is known (solid) and when it is not (dashed). While overall result shown in violet is nearly identical for both fitting routines, amounts of signal and background events are very different. The large width of the signal peak reduces the ability of the fit to distinguish between signal and background contributions in the peak region. Consequently, an overly simplistic description of the background can absorb a fraction of genuine signal events, resulting in what might be called signal leakage.

\begin{figure}[H]
\begin{center}
\includegraphics[width=0.38\textwidth,angle=0]{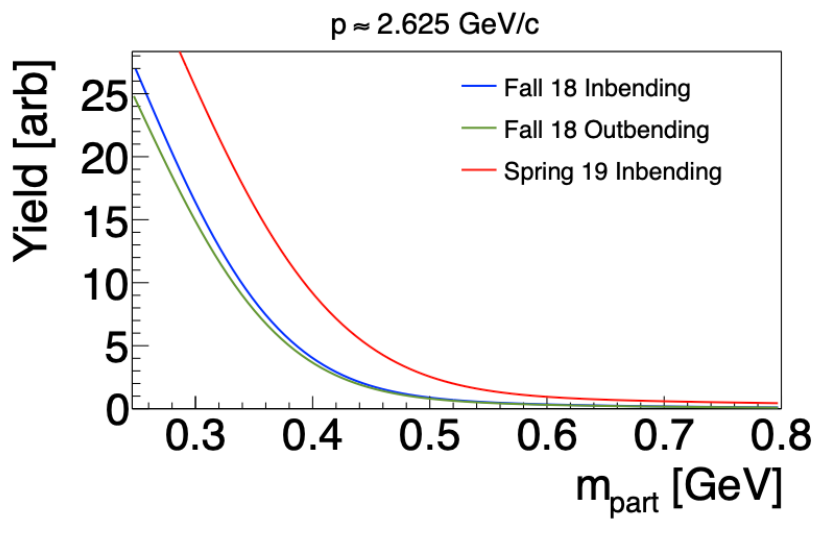}
\end{center}
\caption{Comparison of pion function shape at $p \approx 2.625 $ GeV for data taken in Spring 2019 with the torus magnet in the \emph{inbending} configuration, Fall 2018 with the torus magnet in the \emph{outbending} configuration and Fall 2018 with the torus magnet in the \emph{inbending} configuration.}
\label{PionTailFitComparison}
\end{figure}

This procedure can be repeated for all different run groups and toroidal magnet orientations. A typical comparison plot for a given slice in $p$ looks like Fig.~\ref{PionTailFitComparison}. The shape of the pion tail stays essentially unchanged regardless of  magnet polarity (\emph{inbending}/\emph{outbending}), but may slightly vary depending on run group partly due to variation in calibrations and related detector resolution. 

\section{Error Calculation by Bootstrapping \label{sec::Boot}}

With such a complex multi-step analysis routine, proper error evaluation for yield extraction would always be a concern. Hence, use of a mathematically rigorous process for uncertainty evaluation is highly desirable. We believe that the bootstrap procedure provides high reliability independent on nature of data and fitting procedures employed on previous steps~\cite{Pastore_2019}. In this section we will demonstrate how statistical uncertainties on the fitted distributions for the multi-dimensional sequential sWeight analysis can be estimated using a bootstrap resampling procedure.

Each bootstrap replica is generated by sampling the original event sample with replacement, producing a dataset containing the same number of events as the original\footnote{In a new replica some of the original events might appear several times and some never. Each replica contains different sets of events.}. The complete fitting and sequential weighting procedure is then repeated independently for every replica, allowing the statistical fluctuations of the entire analysis chain to be propagated to the final distributions.

For each bootstrap sample $i$, a "signal/control" histogram $H_i(b)$ is constructed, in this particular case, as a function of the missing mass bin $b$. The bootstrap estimator of the mean distribution, $\bar{H}(b)$, is then given by
\begin{align}
    \bar{H}(b) = \frac{1}{N} \sum^{N}_{i=1} H_i (b). 
\end{align}
where $N$ denotes the number of bootstrap replicas \cite{EfronTibshirani1993}. In this analysis $N \sim 2000$ replicas were used. The statistical uncertainty in each missing-mass bin, $\sigma_H(b)$, is then estimated as
\begin{align} \label{bootstraperr}
    \sigma_H(b) = \sqrt{\frac{1}{N-1} \sum^{N}_{i=1} (H_i (b) - \bar{H}(b))^2}. 
\end{align}
Relative uncertainty on the estimated standard deviation is approximately:
\begin{align}
    \frac{1}{\sqrt{2(N-1)}}. 
\end{align}
The final signal distribution is obtained through a multi-stage fitting and weighting procedure, so analytic propagation of statistical uncertainties is highly impractical, especially since in some cases uncertainties might be asymmetrical and/or non-gaussian. This procedure provides a robust estimate of the statistical uncertainty associated with the fitting method, including effects from fluctuations in the input data.
\begin{figure} [h]
    \centering
    \includegraphics[width=1\linewidth]{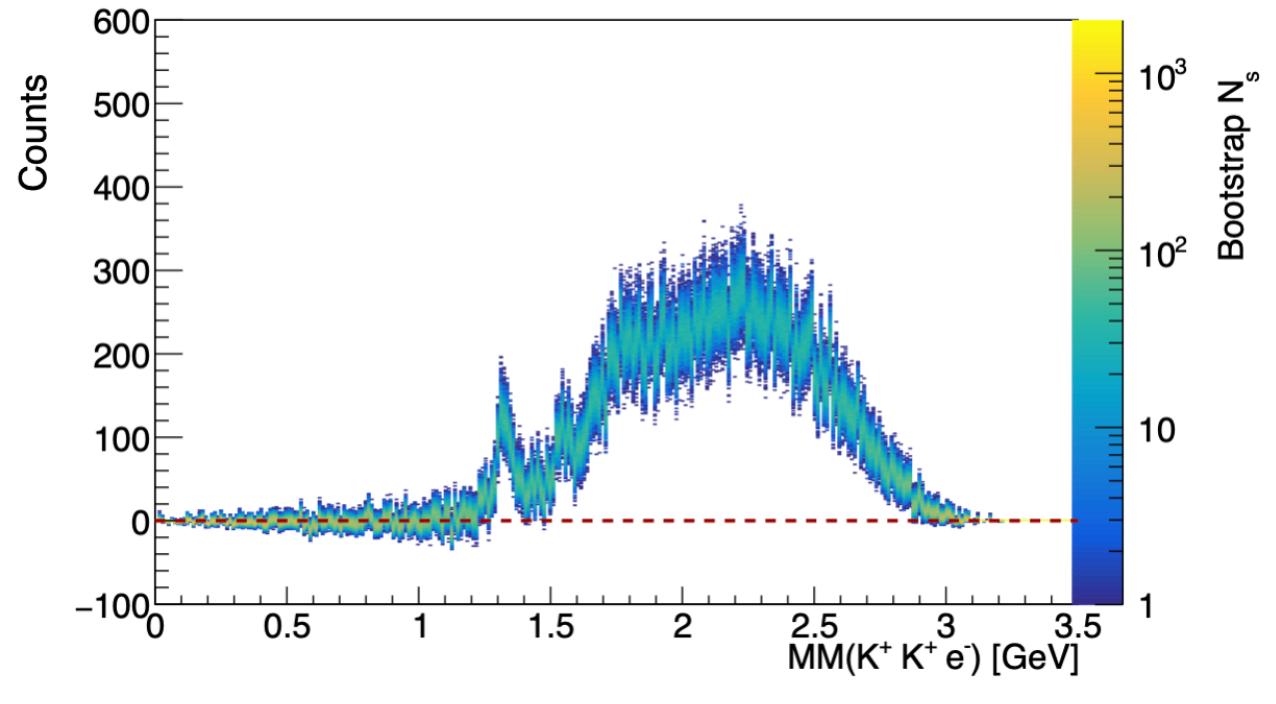}
    \caption{Signal yield for each bootstrap sample per missing mass ($MM(K^+K^+e^-)$) bin, exhibiting the spread of yields for each bootstrap fit for the reaction channel $ep \rightarrow e'K^+K^+X$ where $X$ represents the undetected decay product.}
    \label{fig:Boot}
    
\end{figure}
Figure~\ref{fig:Boot} shows the signal yield obtained from each of the $\sim 2000$ bootstrap replicas as a function of missing mass. The spread of the replicas in each bin provides the statistical uncertainty according to Eq.~\ref{bootstraperr}. 
\begin{figure}[h]
\begin{center}
\includegraphics[width=0.42\textwidth,angle=0]{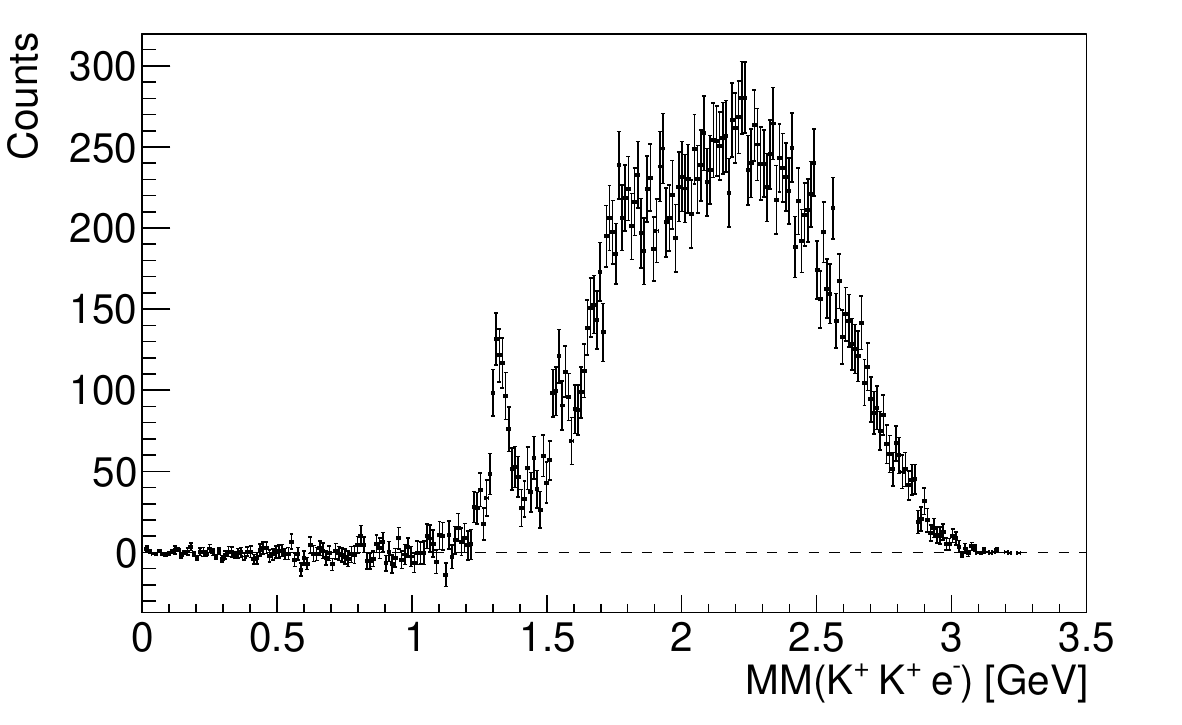}
\end{center}
\caption{Data taken in Fall 2018 with the torus magnet in \emph{inbending} mode. Total signal sWeighted distribution for the missing mass distribution, $MM(K^+K^+e^-)$, with errors extracted with the bootstrap method for the reaction channel $ep \rightarrow e'K^+K^+X$ where $X$ represents the undetected decay product.}
\label{SignalBootstrap}
\end{figure}
The resulting uncertainty band on the final signal distribution is shown in Fig.~\ref{SignalBootstrap}.

\section{Correlation Studies \label{sec::corr}}

To ensure that the method is working reliably under various experimental conditions we performed a synthetic data study, evaluating validity of the method for various mathematical extremes rarely seen in real data. The results of such studies are reported in this section. 
 
sWeights relies on the assumption that the discriminating variables used in the fit are independent of the control variable used for projection. In the limiting case of correlation between the two, the decomposition into separable signal and background components becomes ill-defined.

Nevertheless, the method retains robustness under moderate correlations, although the resulting weights may exhibit increased bias and variance depending on the correlation structure.

Here, we study the impact of correlations in a sequential multidimensional implementation, focusing on correlations among the discriminating observables $p_1$ and $p_2$, as well as correlations between the control variable and the discriminating variables.

\subsection{Correlation between discriminatory variables and the control variable}

In the limiting case where the control variable is completely determined by the discriminatory variables, the information contained in the control variable is already encoded in the fit observables, and the assumptions underlying the sWeight decomposition cease to be valid. Although the sWeight formalism assumes independence between the control variable and the discriminating observables, in practice small residual correlations can be tolerated.

To investigate the effect of such correlations, a simple Monte Carlo model was constructed. The signal PDF of $p_1$ and $p_2$ was constructed a two-dimensional Gaussian, and the background PDF as a two dimensional exponential. In this case, we want to have full control over the correlation between discriminatory variables, as we don't want that to influence the effects of the correlation between the control variable and discriminatory variables. In order to have full control over the correlations between all variables, these correlations were introduced through the covariance matrix of a multivariate Gaussian distribution. The covariance matrix was constructed with $\rho_{xy}=0$, ensuring that $p_1$ and $p_2$ were generated without intrinsic correlations. The covariance matrix was decomposed using a Cholesky decomposition~\cite{golub2013matrix}.

A control variable $z$ was then generated in the signal region according to
\begin{align}
    z = z_0 + \alpha p_1 + \beta p_2 + \epsilon ,
\end{align}
where
\begin{align}
    \epsilon \sim \mathcal{N}(0,\sigma_z^2),
\end{align}
with $z_0=1.2$ and $\alpha, \beta$ as free parameters. The parameter $\alpha$ was varied to introduce increasing correlation between $z$ and the discriminatory variable $x$, while $\beta=0.05$ was fixed. As $\alpha$ increases, the control variable becomes progressively more dependent on the fit observables, thereby violating the independence assumption required by the sWeight method. The values for $z$ in the background region were generated using a two-dimensional exponential distribution, with no injected correlation.

The effect of this correlation was quantified by comparing the true signal distribution in $z$ with the distribution reconstructed using sWeights. Shape distortions were characterized using the metric
\begin{align}
    \chi^2_{\mathrm{shape}}
    =
    \frac{1}{N_{\mathrm{bins}}}
    \sum_{i=1}^{N_{\mathrm{bins}}}
    \frac{\left(A_i-B_i\right)^2}
         {A_i+\varepsilon},
\end{align}
where $A_i$ and $B_i$ denote the contents of bin $i$ in the true and reconstructed signal distributions, respectively, and $\varepsilon$ is a small regularization constant introduced to avoid division by zero.

\begin{figure}[H]
\begin{center}
\includegraphics[width=0.5\textwidth,angle=0]{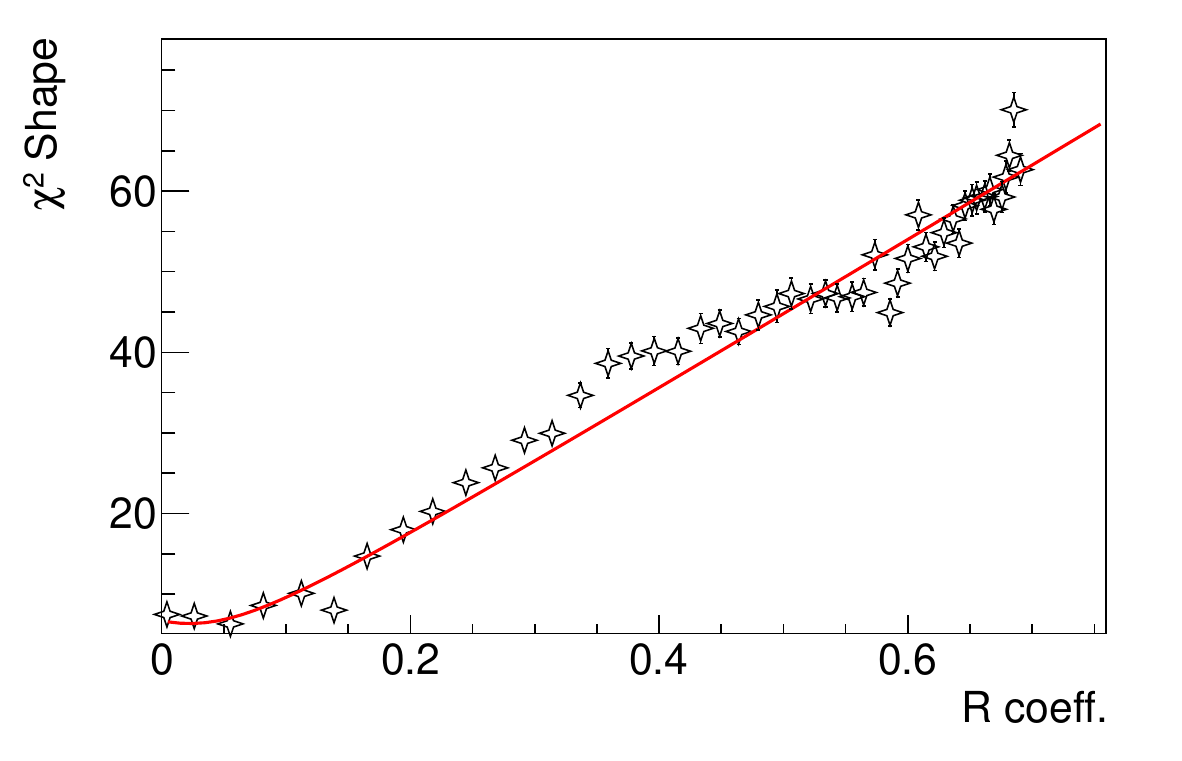}
\end{center}
\caption{Dependence of the shape distortion metric on the Pearson correlation coefficient between the discriminatory variable $x$ and the control variable $z$, fit with a square root quadratic function.}
\label{Chi2Shape}
\end{figure}

A total of $N = 10^6$ events were generated for $50$ distinct $\alpha$ values, equally spaced between $0$ and $2.5$. Figure~\ref{Chi2Shape} shows the dependence of this shape metric on the Pearson correlation coefficient between $x$ and $z$. A clear monotonic increase is observed, indicating that stronger correlations lead to progressively larger distortions of the reconstructed signal shape. This behavior is consistent with expectations, since increasing dependence between the control variable and the discriminatory variables violates one of the central assumptions of the sWeight formalism.

In addition to shape agreement, the quality of the signal yield reconstruction was assessed using the closure quantity
\begin{align}
    C =
    \frac{|S_{\mathrm{true}}-S_{\mathrm{reco}}|}
         {S_{\mathrm{true}}},
\end{align}
where $S_{\mathrm{true}}$ is the generated signal yield and $S_{\mathrm{reco}}$ is the signal yield obtained from the sum of weights.

\begin{figure}[H]
\begin{center}
\includegraphics[width=0.5\textwidth,angle=0]{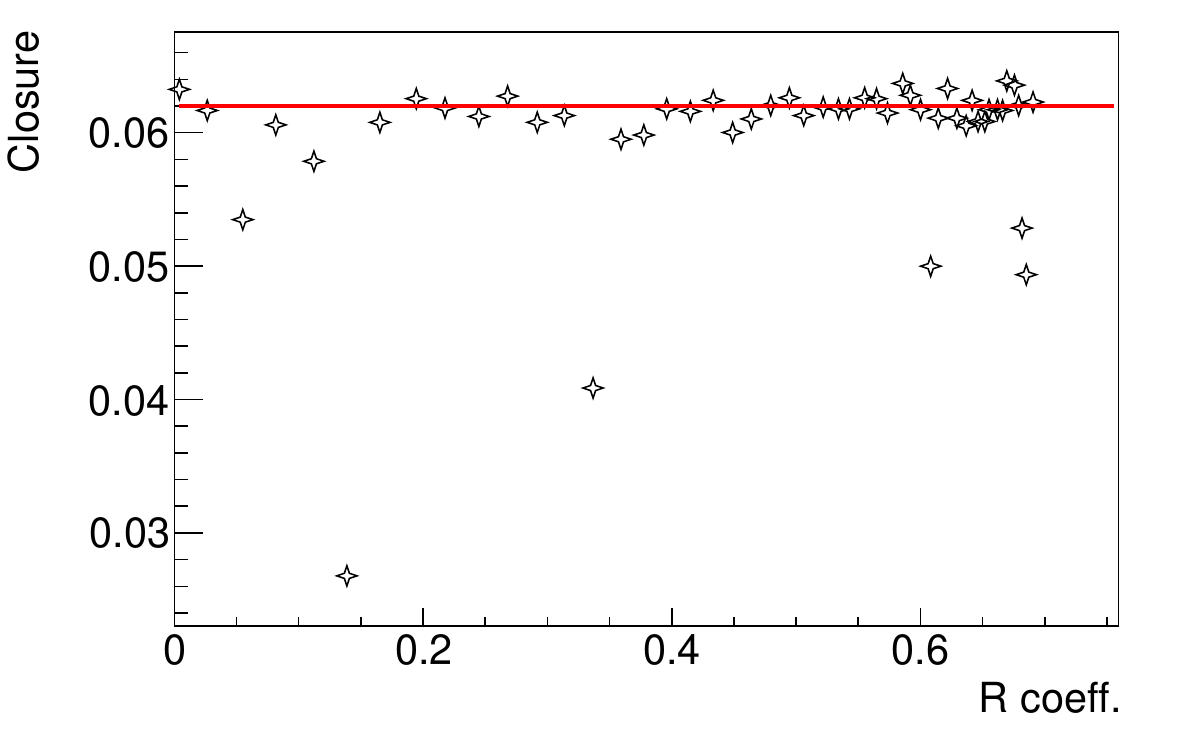}
\end{center}
\caption{Dependence of the closure metric on the Pearson correlation coefficient between $x$ and $z$.}
\label{ClosureMM}
\end{figure}

As shown in Figure~\ref{ClosureMM}, the closure remains largely stable across the investigated correlation range, with only a weak increase observed at larger correlation coefficients. This suggests that the primary impact of violating the independence assumption is not a failure to reproduce the overall signal yield, but rather a distortion of the reconstructed signal shape. Consequently, closure alone is not sufficient to diagnose the effects of correlations; shape-based metrics provide a considerably more sensitive probe of deviations from the assumptions underlying the sWeight method. 

\subsection{Correlation between discriminatory variables}

The control variable is first constructed such that it is uncorrelated with the discriminating observables $p_1$ and $p_2$ in the baseline configuration. The signal and background probability density functions for $p_1$ and $p_2$ are then initially treated as factorizable. To study the impact of residual signal correlations, we introduce a controlled model in which the signal distributions of $p_1$ and $p_2$ exhibit a tunable linear correlation for a fraction $f$ of events. In this model, increasing $f$  enhances the contribution of correlated signal components, while the background remains unchanged.

By varying $f$, the strength of the induced correlation is systematically scanned, and its effect on the closure of the sequential sWeighting procedure is evaluated. In this case, the effect of the correlation on the yield is more significant than shape distortion. The signal and background PDFs are chosen to be separable.

\begin{figure}[H]
    \centering
    \includegraphics[width=1\linewidth]{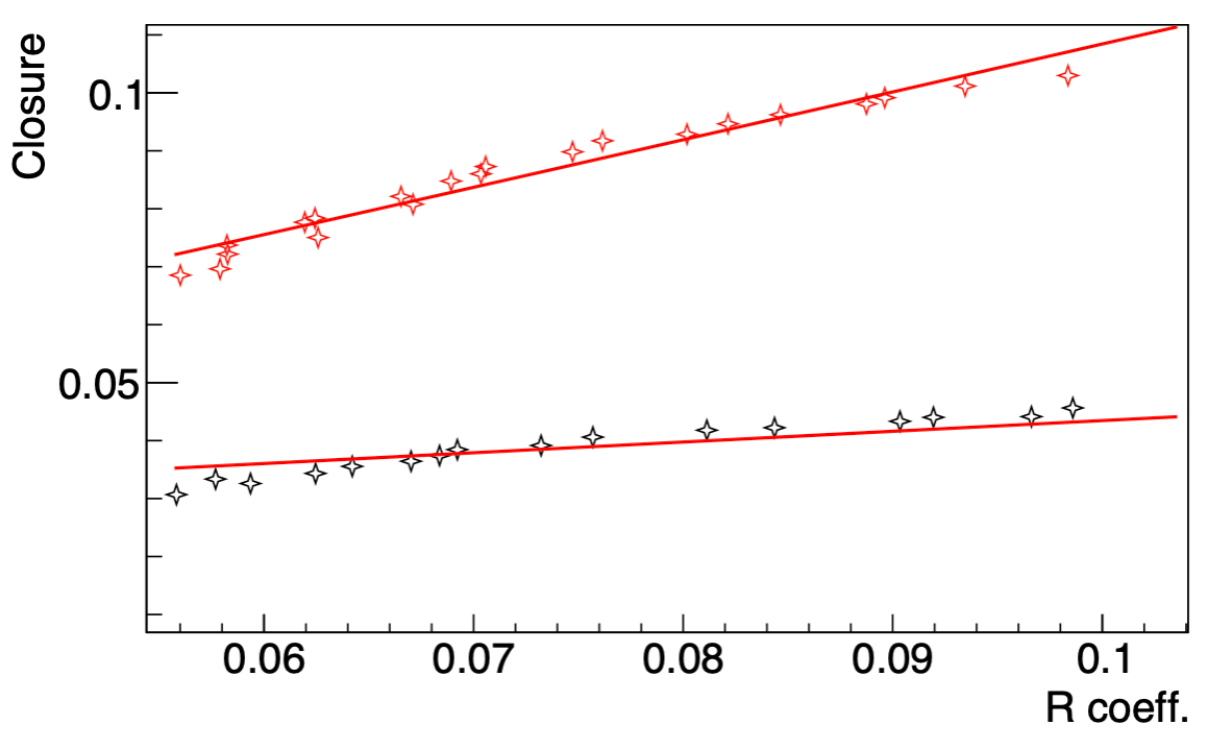}
    \caption{Closure on the y-axis and Pearson correlation coefficient on the x-axis, fitted with a function of the form $Ax-\frac{B}{(x-C)^2}+D$. Black: separable signal and background PDFs. Red: Significantly overlapping signal and background PDFs.}
    \label{fig:KaonCorrClosure}
\end{figure}

Fig. \ref{fig:KaonCorrClosure} shows the effect of unmodelled linear correlation between $p_1$ and $p_2$ in the signal PDF, while the fit assumes factorization.
Overall, correlation between any of the components of the fitting procedure should be handled carefully. A moderate correlation does not automatically render the method invalid, however, the effects of it should carefully be explored with Monte Carlo methods on a case by case basis. Another important factor to consider is the separability of signal and background PDFs in the control variable. If the shapes of the PDFs are too similar, it can also introduce an error in the yield estimation. Fig. \ref{fig:KaonCorrClosure} shows the effect of the separability of the PDFs on closure, in red data points.
At a moderate level of linear correlation, corresponding to a Pearson coefficient of 0.1, the uncertainty on the extracted yield is found to be approximately 4.5 percent. The magnitude of the Pearson correlation coefficient between kaon mass and missing mass for the CLAS12 data samples is in the range [0.03, 0.04]. The correlation between kaon masses is in the range [0.008, 0.065]. The observed fluctuations in the closure are attributed to imperfections in the construction of the covariance matrix used in the weighting procedure.

\section{Results}
\subsection{Final Model}
The fitting model that is the most physics-informed gives the best results when it comes to modeling the kaon mass spectrum. This model utilizes 100 MeV wide Gaussian sampling to capture the momentum dependence of signal and background intensity, and contains the pion and proton tails extracted in Section~\ref{chap::PiPr}. The shape of the signal is given by $S(m,p)$ in Eq.~\ref{Gaussians}, and its momentum dependent width by Eq.~\ref{GaussiansW}. The crossings have fixed width, as we determined that due to the narrow range of momentum they cover, the momentum dependence of their width is negligible. The parameter $d$ of these crossings is extracted using the procedure described in Section~\ref{chap::Cross}, and fixed.

\subsection{Final Distributions}

\begin{figure}[h]
\begin{center}
\includegraphics[width=0.45\textwidth,angle=0]{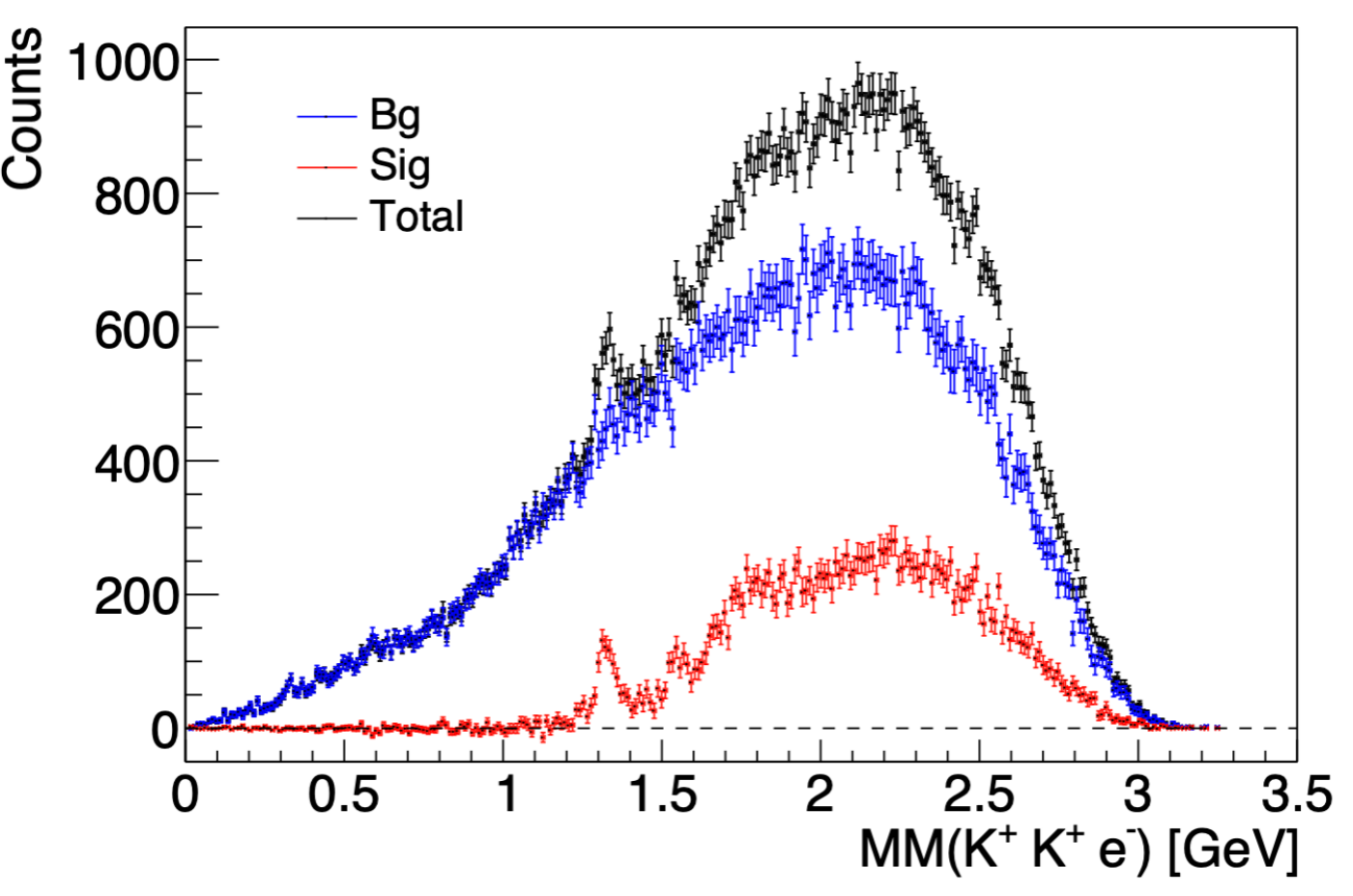}
\end{center}
\caption{Data taken in Fall 2018 with the torus magnet in \emph{inbending} mode,  showing the missing mass distribution, $MM(K^+K^+e^-)$ for the reaction channel $ep \rightarrow e'K^+K^+X$ where $X$ represents the undetected decay product. Total distribution is given in black, sWeighted signal and background distributions given in red and blue respectively.}
\label{FinalFullSignalsFD}
\end{figure}

The final total, signal weighted and background weighted distributions for data taken in Fall 2018 with the torus magnet in the \emph{inbending} configuration is shown for the FD case in Fig.~\ref{FinalFullSignalsFD}. The signal distribution slightly fluctuates around zero below threshold, the ground state cascade, $M_{\Xi}=1.32$~GeV, where no physical event can exist. The background shows smooth monotonic behavior with all the $\Xi^{(*)}$ related peaks staying in signal. The final distributions of the other datasets used in this study are given in Appendix A.

\begin{figure}[h]
\begin{center}
\includegraphics[width=0.45\textwidth,angle=0]{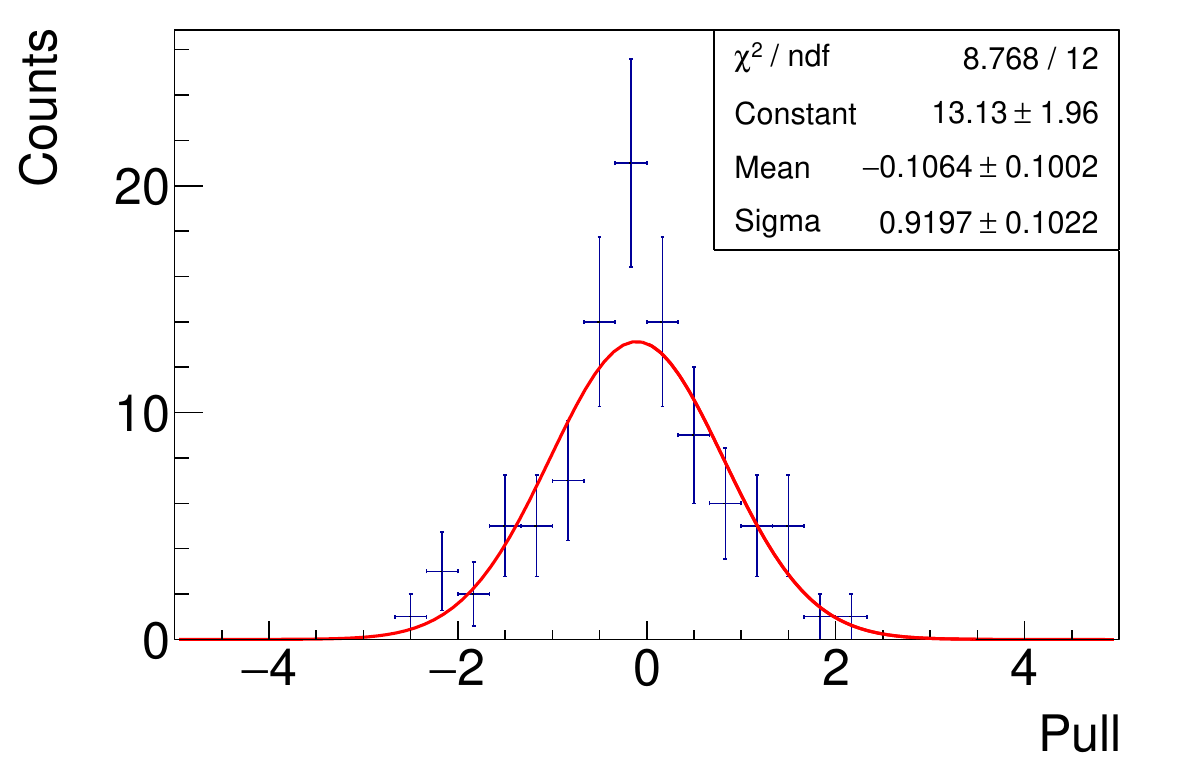}
\end{center}
\caption{Data taken in Spring 2019 with the torus magnet in the \emph{inbending} configuration. Pull distribution for 2D model and data distributions projected along the kaon mass. Gaussian fitted to the pull distribution in red, parameters of the fit given in the legend.}
\label{PullsS19K1}
\end{figure}

An example of the pull distribution for 2D model and data distributions projected along the kaon mass is given in Fig.~\ref{PullsS19K1}. As expected for a good fit, the pull distribution is centered near 0 with a width close to 1.

\begin{figure*}[t]
\centering
\includegraphics[width=0.4\textwidth,angle=0]{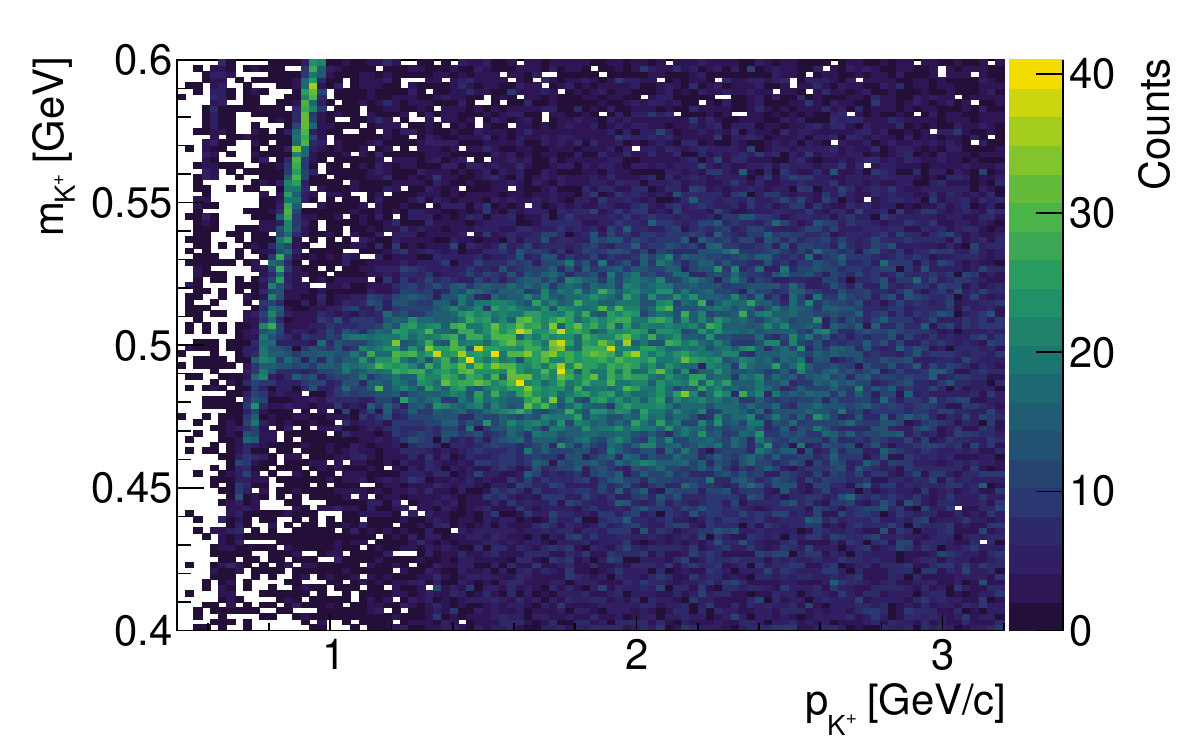}
\includegraphics[width=0.4\textwidth,angle=0]{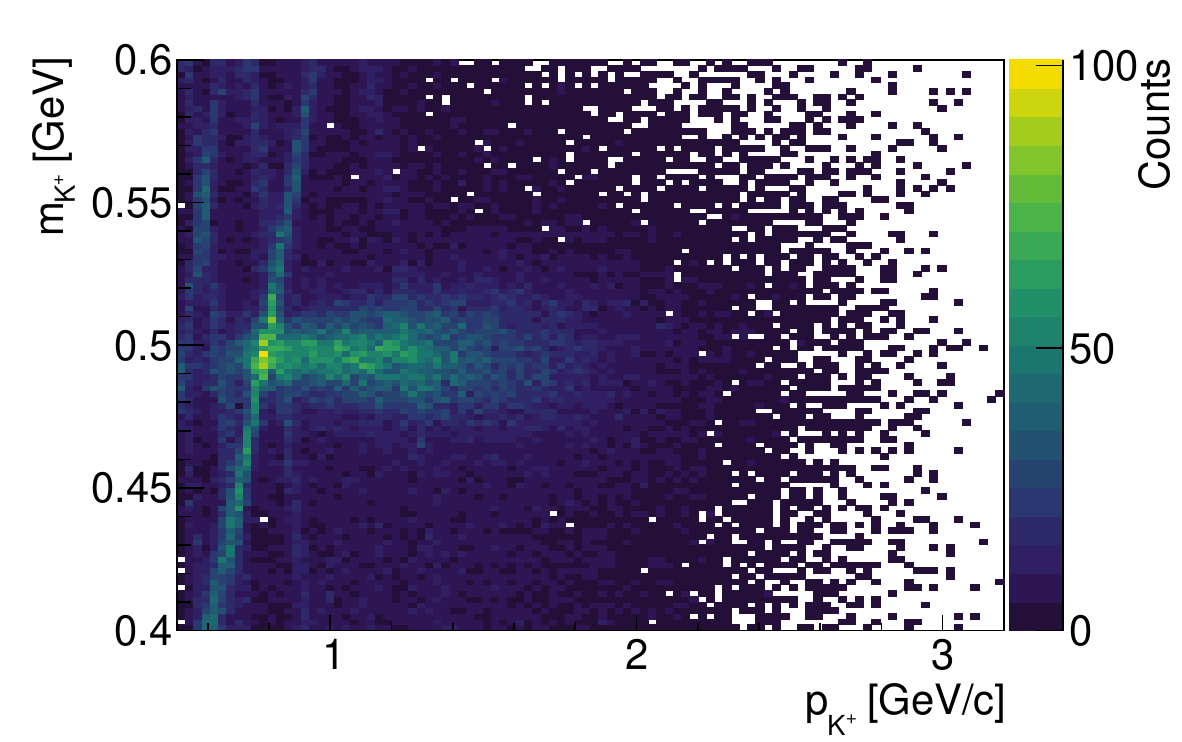}
\includegraphics[width=0.4\textwidth,angle=0]{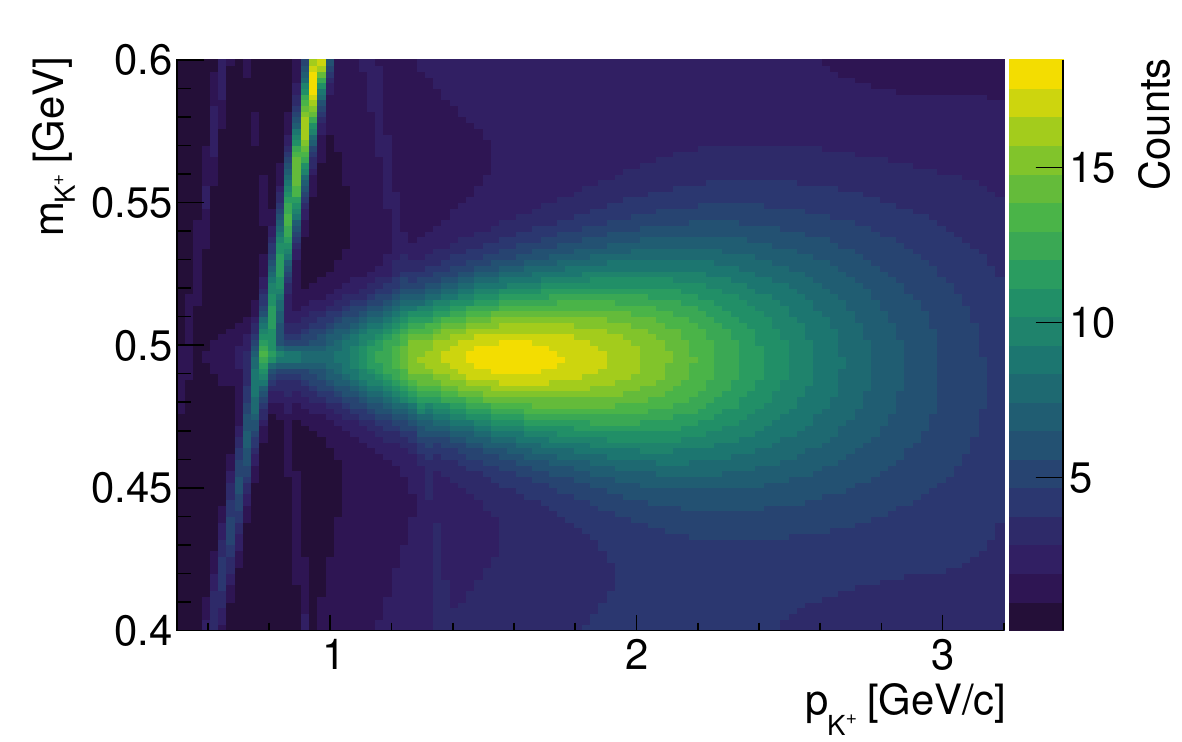}
\includegraphics[width=0.4\textwidth,angle=0]{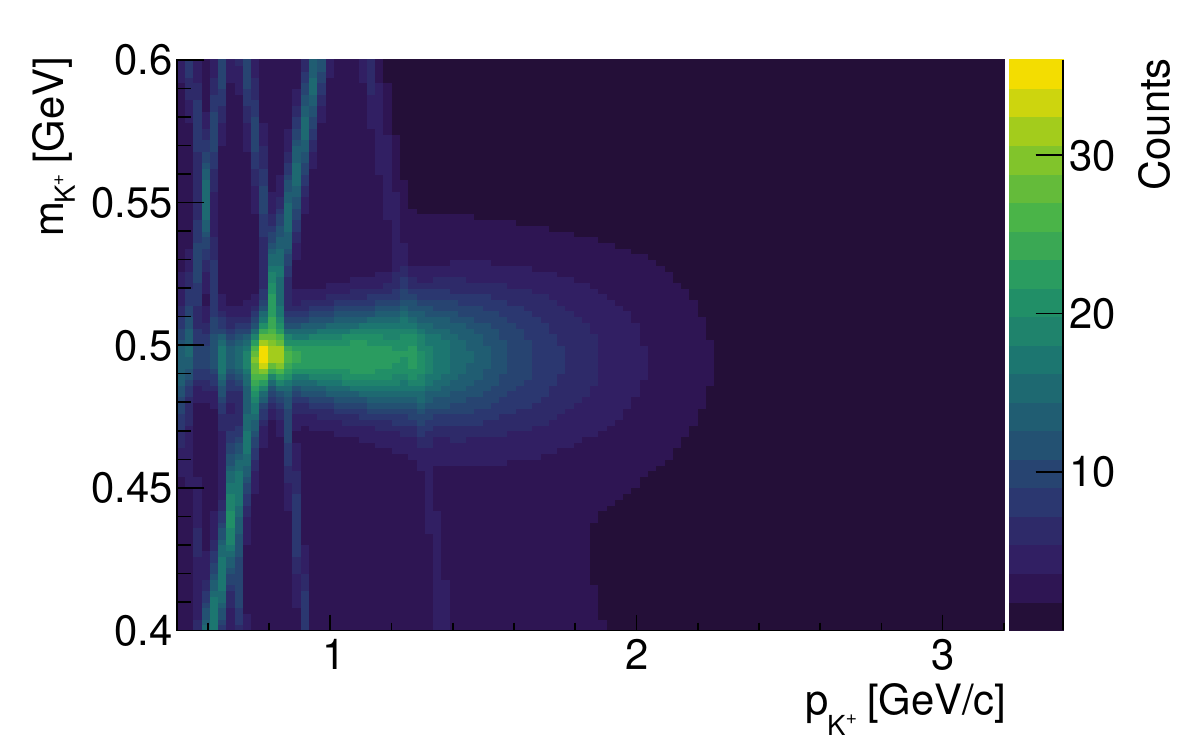}
\caption{Data taken in Spring 2019 with the torus magnet in the \emph{inbending} configuration. \underline{Top left}: kaon 1 mass plotted against its momentum. \underline{Top right}: kaon 2 mass plotted against its momentum. \underline{Bottom left}: Two dimensional log-likelihood fit of kaon 1. \underline{Bottom right}: Two dimensional log-likelihood fit of kaon 2.}
\label{2D plots of K}
\end{figure*}

The resulting fitted model distributions alongside real data for kaon 1 and kaon 2 is shown in Fig.~\ref{2D plots of K}. 

\begin{figure} [h!]
\centering
\includegraphics[width=0.76\linewidth]{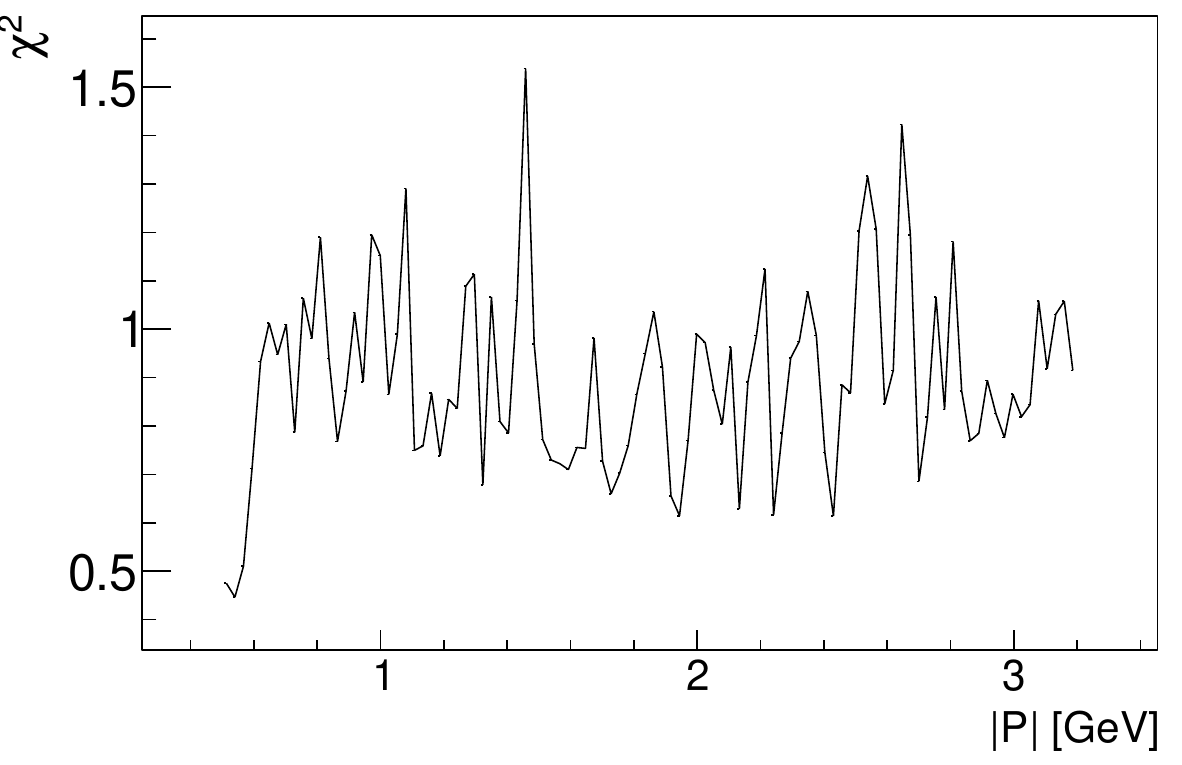}
\includegraphics[width=0.76\linewidth]{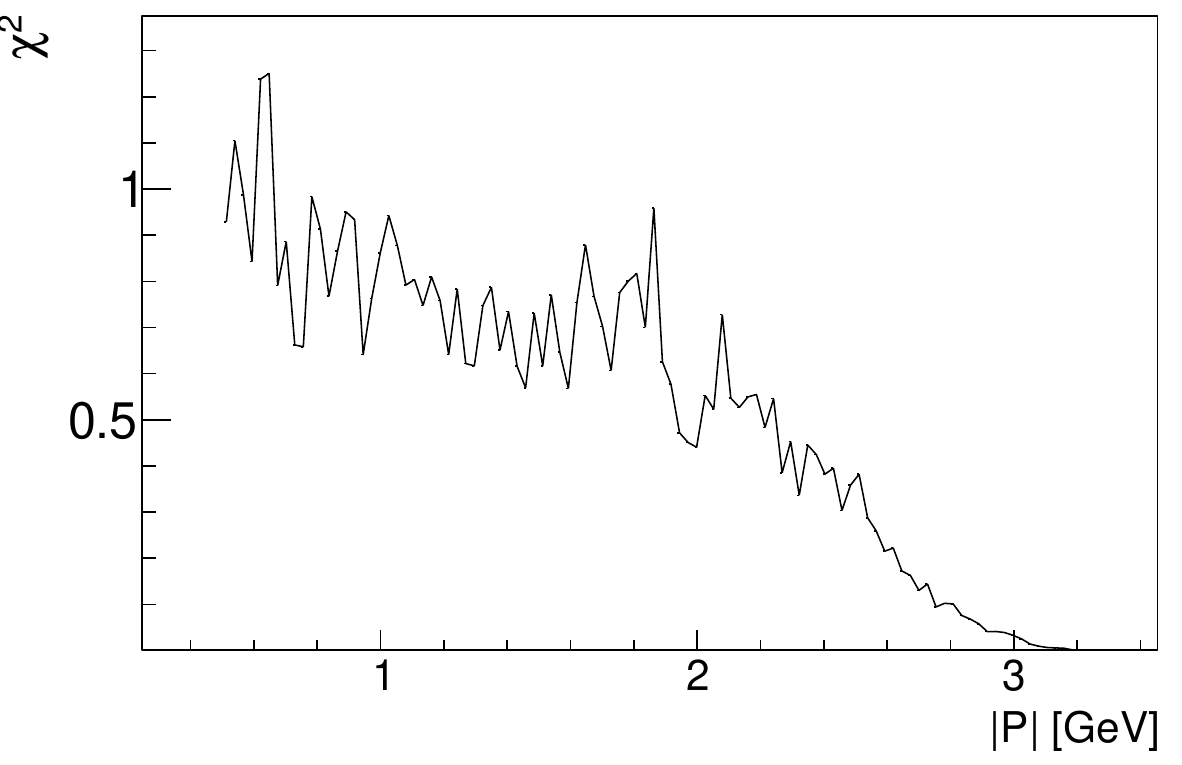}
\caption{Data taken in Spring 2019 with the torus magnet in the \emph{inbending} configuration. \underline{Top}: $\chi^2$ of kaon 1, \underline{Bottom}:  $\chi^2$ of kaon 2.}
\label{fig: Chi2s}
\end{figure}

The $\chi^2$ distributions for both kaon 1 and kaon 2  as a function of momentum are shown in Fig.~\ref{fig: Chi2s} demonstrating $\chi^2\sim 1$ per dof.

\begin{figure} [H]
\centering
\includegraphics[width=1.1\linewidth]{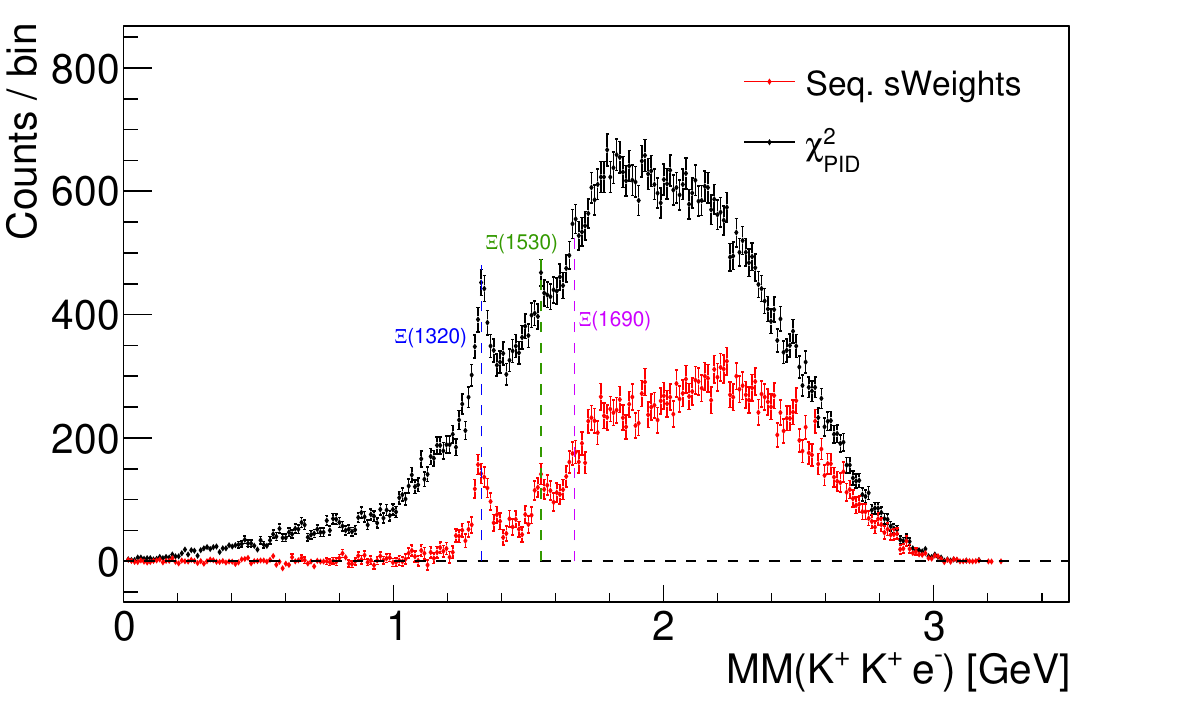}
\caption{Data taken in Fall 2018 with the torus magnet in the \emph{inbending} configuration, showing the missing mass distribution, $MM(K^+K^+e^-)$ for the reaction channel $ep \rightarrow e'K^+K^+X$ where $X$ represents the undetected decay product. The sequential sWeights signal distribution is shown in red and the $\chi^2_{PID}<3$ distribution in black. The nominal masses of the $\Xi(1320)$, $\Xi(1530)$, and $\Xi(1690)$ are indicated by blue, green, and violet dashed lines, respectively.}
\label{fig: Chi2Comp}
\end{figure}

The comparison of the final missing-mass distributions obtained using the sequential sWeights method and using only $\chi^2_{PID}<3$ requirements on all final-state particles is shown in Fig. \ref{fig: Chi2Comp}. The dashed lines indicate the nominal masses of the $\Xi(1320)$, $\Xi(1530)$, and $\Xi(1690)$ states, allowing the peak positions to be compared between the two methods. As can be seen, the signal distribution obtained using the sequential sWeights method fluctuates around zero below the ground-state cascade mass, as expected. In contrast, the distribution obtained using the $\chi^2_{PID}<3$ requirements contains a substantial number of events below the ground-state mass, indicating that subthreshold background remains in the selected signal distribution.

\begin{figure} [H]
\centering
\includegraphics[width=1.0\linewidth]{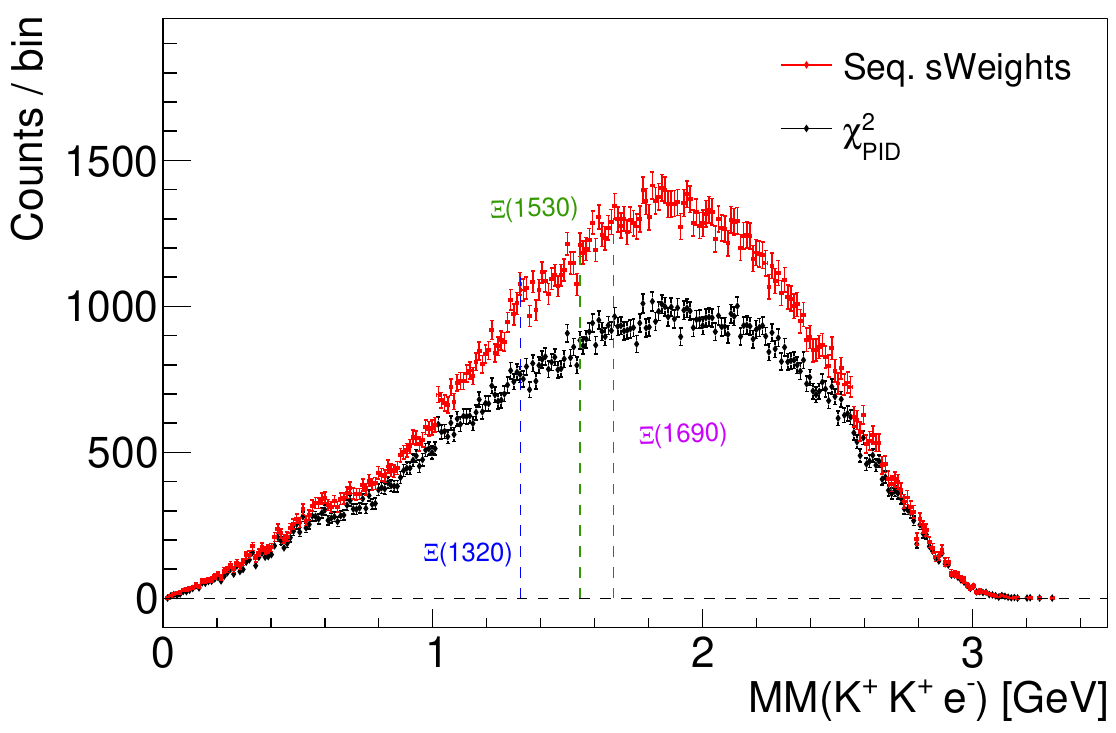}
\caption{Data taken in Fall 2018 with the torus magnet in the \emph{inbending} configuration. Sequential sWeights background distribution in red and  $\chi^2_{PID}<3$ cut distribution in black, with bin width 11.7 MeV. $\Xi(1320)$, $\Xi(1530)$, $\Xi(1690)$ masses marked in blue, green and violet dashed lines respectively.}
\label{fig: Chi2CompBg}
\end{figure}

Signal leakage into the background distribution can, in principle, manifest itself as non-uniform structures at the locations of the expected signal states. This is particularly relevant in the present analysis, where the signals of interest are expected to appear as peaks or dips in the missing-mass distributions. The final background distributions obtained using the two methods are shown in Fig. \ref{fig: Chi2CompBg}. The dashed lines again indicate the nominal masses of the $\Xi(1320)$, $\Xi(1530)$, and $\Xi(1690)$ states, marking the locations where non-uniform structures would be expected in the presence of signal leakage. Neither method shows a statistically significant fluctuation at these locations, indicating no evidence of appreciable signal leakage into the background distributions.

\section{Conclusion}
A novel multi-dimensional consecutive background suppression technique was developed. Its performance and validity were verified with the two-dimensional sWeight procedure applied to kaon identification in CLAS12 cascade-baryon analyses. This technique reduces the complexity of the original 4D problem by factorizing the joint distribution into sequential single-particle components, thereby avoiding the need for a full 4D unbinned fit, diminishing computational cost, while retaining the statistical advantages of an event-by-event weighting approach. The method provides effective signal-background separation while retaining the statistical advantages of an event-by-event weighting approach. A clear demonstration of its performance is shown in Fig.~\ref{SignalBootstrap}, where both the ground-state and first excited cascade signals are preserved while the sub-threshold background is eliminated completely. The approach relies on approximate factorization assumptions and has some sensitivity to residual correlations between discriminatory variables, which are treated via iterative fitting and bootstrap procedure discussed in previous sections.

The method was applied to multiple CLAS12 RGA datasets, consisting of the Spring 2019 dataset with the torus magnet in \emph{inbending} mode, Fall 2018 dataset with the torus magnet in \emph{inbending} mode, and Fall 2018 dataset with the torus magnet in \emph{outbending} mode, under very different kinematical conditions (electrons in FD or FT). The consistent reconstruction of the cascade signal across these datasets indicates that the procedure is stable under varying detector and beam conditions and can be readily adapted to similar analyses. The results of the validation study can be seen in Appendix A.

A dedicated toy Monte Carlo study was also performed to investigate the impact of correlations between the discriminatory variables used in the likelihood fit and the control variable reconstructed with sWeights. While signal-yield closure was found to be relatively insensitive to moderate correlations, the reconstructed signal shape exhibited increasing distortions as the correlation strength increased. It is recommended that this study be performed for analyses utilizing this technique, as the effect of correlations are best treated and quantified on a case by case basis.

The improved particle identification achieved with this approach is expected to benefit future measurements of strange-hadron production, where efficient background suppression is essential for isolating low-yield channels. The framework presented here can be extended to additional datasets and reaction channels and provides a practical foundation for future applications of multidimensional sequential sWeight techniques within the CLAS12 physics program or any other experiment.

\section{Acknowledgments}
This work has been supported by the U.K. STFC ( ST/Y000285/1, ST/T002077/1, 57071/1, 50727/1) grants, and by an STFC PhD studentship.
%************
%\section{Acknowledgement}

\printbibliography

@article{Pivk:2005splot,
author        = {Pivk, Muriel and Le Diberder, François R.},
title         = {sPlot: a statistical tool to unfold data distributions},
journal       = {Nucl. Instrum. Meth. A},
volume        = {555},
pages         = {356--369},
year          = {2005},
doi           = {10.1016/j.nima.2005.08.106},
eprint        = {physics/0402083},
archivePrefix = {arXiv},
primaryClass  = {physics.data-an}
}

@article{CLAS12,
author  = {Burkert, V. D. and others},
title   = {The CLAS12 Spectrometer at Jefferson Laboratory},
doi     = {10.1016/j.nima.2020.163419},
journal = {Nuclear Instruments and Methods in Physics Research Section A: Accelerators, Spectrometers, Detectors and Associated Equipment},
volume  = {959},
pages   = {163419},
year    = {2020}
}

@article{Aaij_2015,
title     = {Observation of $J/\psi,p$ Resonances Consistent with Pentaquark States in $\Lambda_b^0 \rightarrow J/\psi,K^-p$ Decays},
volume    = {115},
ISSN      = {1079-7114},
DOI       = {10.1103/physrevlett.115.072001},
number    = {7},
journal   = {Physical Review Letters},
publisher = {American Physical Society (APS)},
author    = {Aaij, R. and others},
year      = {2015},
month     = {Aug}
}

@article{FT,
author  = {Acker, A. and others},
year    = {2020},
month   = {04},
pages   = {163475},
title   = {The CLAS12 Forward Tagger},
volume  = {959},
journal = {Nuclear Instruments and Methods in Physics Research Section A: Accelerators, Spectrometers, Detectors and Associated Equipment},
doi     = {10.1016/j.nima.2020.163475}
}

@article{DC,
author  = {Mestayer, M. D. and Adhikari, K. and Bennett, R. P. and Bueltmann, S. and Chetry, T. and Christo, S. B. and Cook, M. and Cuevas, R. C. and Dodge, G. E. and El Fassi, L. and Forest, T. A. and Jacobs, G. and Hartlove, T. and Hayward, T. B. and Kabir, L. and Kuhn, S. E. and McNulty, D. and Nanda, S. and Newton, J. and Taylor, W. M. and Weinstein, L. B. and Ziegler, V.},
year    = {2020},
title   = {The CLAS12 drift chamber system},
volume  = {959},
journal = {Nuclear Instruments and Methods in Physics Research Section A: Accelerators, Spectrometers, Detectors and Associated Equipment},
doi     = {10.1016/j.nima.2020.163518}
}

@article{FTOF,
author  = {Carman, D. S. and Clark, L. and De Vita, R. and Fedotov, G. and Gothe, R. W. and Hollis, G. and Miller, B. and Phelps, E. and Tian, Y. and Trivedi, A. and Wiggins, C.},
year    = {2020},
title   = {The CLAS12 Forward Time-of-Flight system},
volume  = {959},
journal = {Nuclear Instruments and Methods in Physics Research Section A: Accelerators, Spectrometers, Detectors and Associated Equipment},
doi     = {10.1016/j.nima.2020.163629}
}

@article{CTOF,
author  = {Carman, D. S. and Asryan, G. and Baturin, V. and Clark, L. and De Vita, R. and Kim, W. and Miller, B. and Wiggins, C.},
year    = {2020},
title   = {The CLAS12 Central Time-of-Flight system},
volume  = {960},
journal = {Nuclear Instruments and Methods in Physics Research Section A: Accelerators, Spectrometers, Detectors and Associated Equipment},
doi     = {10.1016/j.nima.2020.163626}
}

@article{CND,
author  = {Chatagnon, P. and Bettane, J. and Hoballah, M. and Hull, G. and Imre, M. and Marchand, D. and Mathon, B. and Murdoch, G. and Naidoo, P. and Niccolai, S. and Price, K. and Sokhan, D. and Wang, R.},
year    = {2020},
title   = {The CLAS12 Central Neutron Detector},
volume  = {960},
journal = {Nuclear Instruments and Methods in Physics Research Section A: Accelerators, Spectrometers, Detectors and Associated Equipment},
doi     = {10.1016/j.nima.2020.163441}
}

@article{SVT,
author  = {Antonioli, M. A. and Baltzell, Nathan and Boyarinov, S. and Bonneau, P. and Christo, S. and Cuevas, Chris and Defurne, M. and Derylo, G. and Elouadrhiri, L. and Eng, B. and Ewing, T. and Gilfoyle, Gerard and Gotra, Yuri and Leffel, M. and Mandal, S. and Marzolf, B. and Mcmullen, Marc and Merkin, Mikhail and Miller, R. and Ziegler, Veronique},
year    = {2020},
month   = {02},
pages   = {163701},
title   = {The CLAS12 Silicon Vertex Tracker},
volume  = {962},
journal = {Nuclear Instruments and Methods in Physics Research Section A: Accelerators, Spectrometers, Detectors and Associated Equipment},
doi     = {10.1016/j.nima.2020.163701}
}

@article{BAND,
author  = {Segarra, Elisa and Hauenstein, F. and Schmidt, A. and Beck, A. and Beck, S. and Cruz Torres, Reynier and Denniston, A. and Hrnjic, A. and Kutz, T. and Nambrath, A. and Pybus, Jackson and Pryce, K. and Fogler, C. and Hartlove, T. and Weinstein, Lawrence and Vega, J. and Ungerer, M. and Hakobyan, Hayarpi and Brooks, William and Hen, O.},
year    = {2020},
month   = {04},
title   = {The CLAS12 Backward Angle Neutron Detector (BAND)}
}

@article{splots,
author        = {Pivk, Muriel and Le Diberder, Francois R.},
title         = {sPlot: A Statistical Tool to Unfold Data Distributions},
eprint        = {physics/0402083},
archivePrefix = {arXiv},
reportNumber  = {LAL-04-07},
doi           = {10.1016/j.nima.2005.08.106},
journal       = {Nucl. Instrum. Meth. A},
volume        = {555},
pages         = {356--369},
year          = {2005}
}

@article{Hayrapetyan_2026,
title     = {First Evidence for Mixing-Induced CP Violation in $B_s^0 \rightarrow J/\psi,\phi(1020)$ Decays in pp Collisions at $\sqrt{s}=13$ TeV},
volume    = {136},
ISSN      = {1079-7114},
DOI       = {10.1103/ctqx-vxrj},
number    = {18},
journal   = {Physical Review Letters},
publisher = {American Physical Society (APS)},
author    = {Hayrapetyan, A. and others},
year      = {2026},
month     = {May}
}

@article{Ablikim_2025,
title     = {Observation of $\Lambda_c^+\rightarrow\Lambda a_0(980)^+$ and Evidence for $\Sigma(1380)^+$ in $\Lambda_c^+\rightarrow\Lambda\pi^+\eta$},
volume    = {134},
ISSN      = {1079-7114},
DOI       = {10.1103/physrevlett.134.021901},
number    = {2},
journal   = {Physical Review Letters},
publisher = {American Physical Society (APS)},
author    = {Ablikim, M. and others},
year      = {2025},
month     = {Jan}
}

@phdthesis{wreo32365,
author    = {Nicol, Matthew William},
title     = {Exploring the Strong Interaction through Electroproduction of Exotic Particles},
school    = {University of York},
year      = {2023},
month     = {2},
note      = {PhD thesis}
}

@article{Pastore_2019,
title     = {An Introduction to Bootstrap for Nuclear Physics},
volume    = {46},
ISSN      = {1361-6471},
DOI       = {10.1088/1361-6471/ab00ad},
number    = {5},
journal   = {Journal of Physics G: Nuclear and Particle Physics},
publisher = {IOP Publishing},
author    = {Pastore, A.},
year      = {2019},
month     = {Apr},
pages     = {052001}
}

@book{golub2013matrix,
title     = {Matrix Computations},
author    = {Golub, Gene H. and Van Loan, Charles F.},
edition   = {4},
year      = {2013},
publisher = {Johns Hopkins University Press}
}

@book{EfronTibshirani1993,
author    = {Efron, Bradley and Tibshirani, Robert J.},
title     = {An Introduction to the Bootstrap},
publisher = {Chapman \& Hall/CRC},
year      = {1993},
address   = {New York}
}

@article{Ziegler2020CLAS12,
author  = {Ziegler, V. and Baltzell, N. A. and Bossu, F. and Carman, D. S. and Chatagnon, P. and Contalbrigo, M. and De Vita, R. and Defurne, M. and Gavalian, G. and Gilfoyle, G. P. and Glazier, D. I. and Gotra, Y. and Gyurjyan, V. and Harrison, N. and Heddle, D. and Hobart, A. and Joosten, S. and Kim, A. and Markov, N. and Mancilla, S. and Mestayer, M. D. and Newton, J. and Phelps, W. and Niccolai, S. and Sokhan, D. and Ungaro, M.},
title   = {The CLAS12 software framework and event reconstruction},
journal = {Nuclear Instruments and Methods in Physics Research Section A: Accelerators, Spectrometers, Detectors and Associated Equipment},
volume  = {959},
pages   = {163472},
year    = {2020},
doi     = {10.1016/j.nima.2020.163472}
}

\clearpage
\onecolumn

\appendix

\section{Validation of Additional Datasets}

\begin{figure}[H]
\begin{center}
\includegraphics[width=0.6\textwidth, angle=0]{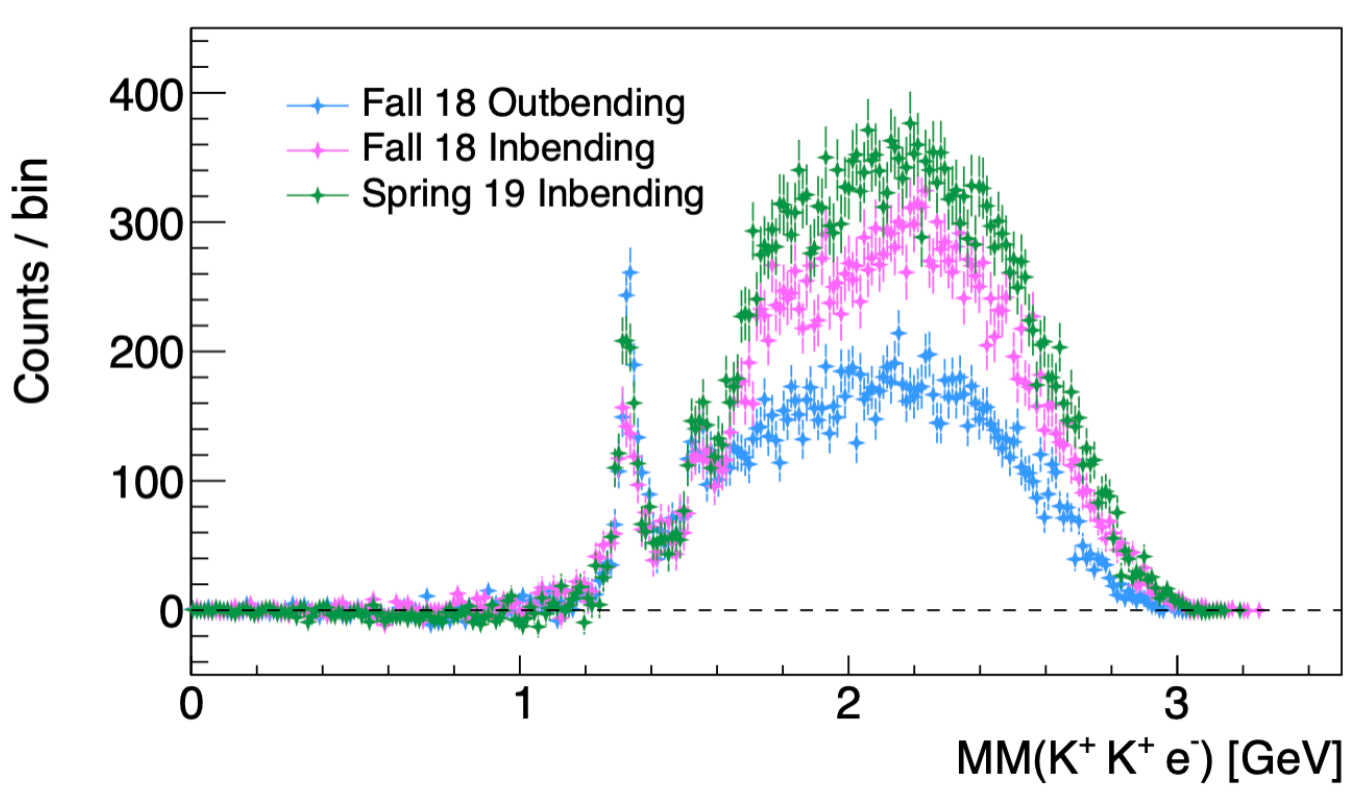}
\includegraphics[width=0.6\textwidth, trim=0.1cm 0cm 0.1cm 0cm,
clip, angle=0]{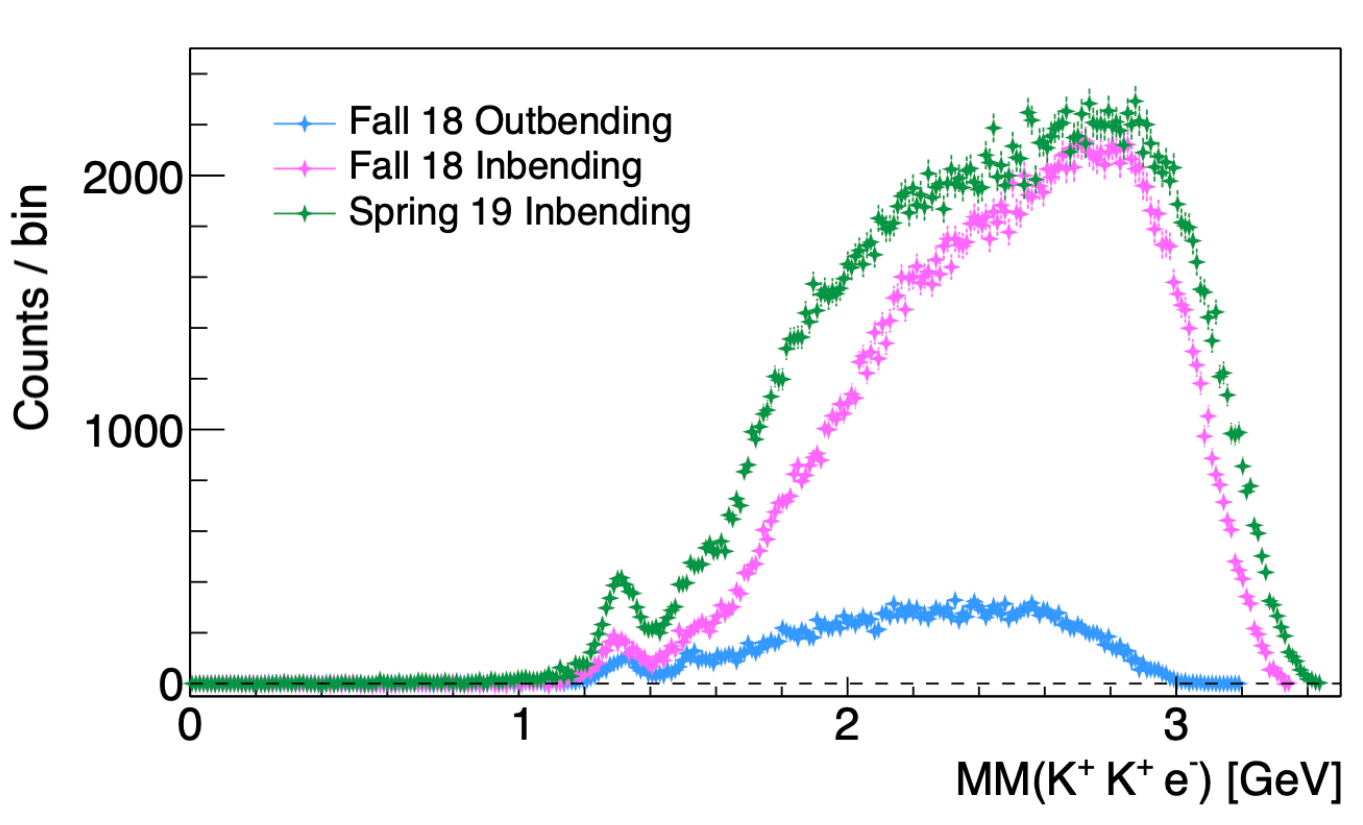}
\end{center}
\caption{Final sWeighted signal distributions. \underline{Top}: Final plots for electron in FD. Data taken in Fall 2018  with the torus magnet in the \emph{outbending} configuration in blue, Fall 2018  with the torus magnet in the \emph{inbending} configuration in pink and Spring 2019  with the torus magnet in the \emph{inbending} configuration in green. (note the difference between the \emph{inbending} and \emph{outbending}). \underline{Bottom}: Final plots for electron in FT. Data taken in Fall 2018  with the torus magnet in the \emph{outbending} configuration in blue, Fall 2018  with the torus magnet in the \emph{inbending} configuration in pink and Spring 2019  with the torus magnet in the \emph{inbending} configuration in green.}
\label{FinalFullSignalsFDFT}
\end{figure}

The final sWeighted signal distributions on data taken in Spring 2019 with the torus magnet in the \emph{inbending} configuration, Fall 2018  with the torus magnet in the \emph{inbending} configuration and Fall 2018 with the torus magnet in the \emph{outbending} configuration is shown for the FD and FT cases in Fig.~\ref{FinalFullSignalsFDFT}. The quality of the weighting proves impressive, with the signal distribution slightly fluctuating around zero before the ground state cascade peak, for both the FD and FT cases. No additional normalization was imposed on either figure. The difference in statistics is purely experimental (length of the beamtime, triggers, \emph{inbending} vs \emph{outbending} acceptance).

\begin{figure}[h]
\begin{center}
\includegraphics[width=0.42\textwidth,angle=0]{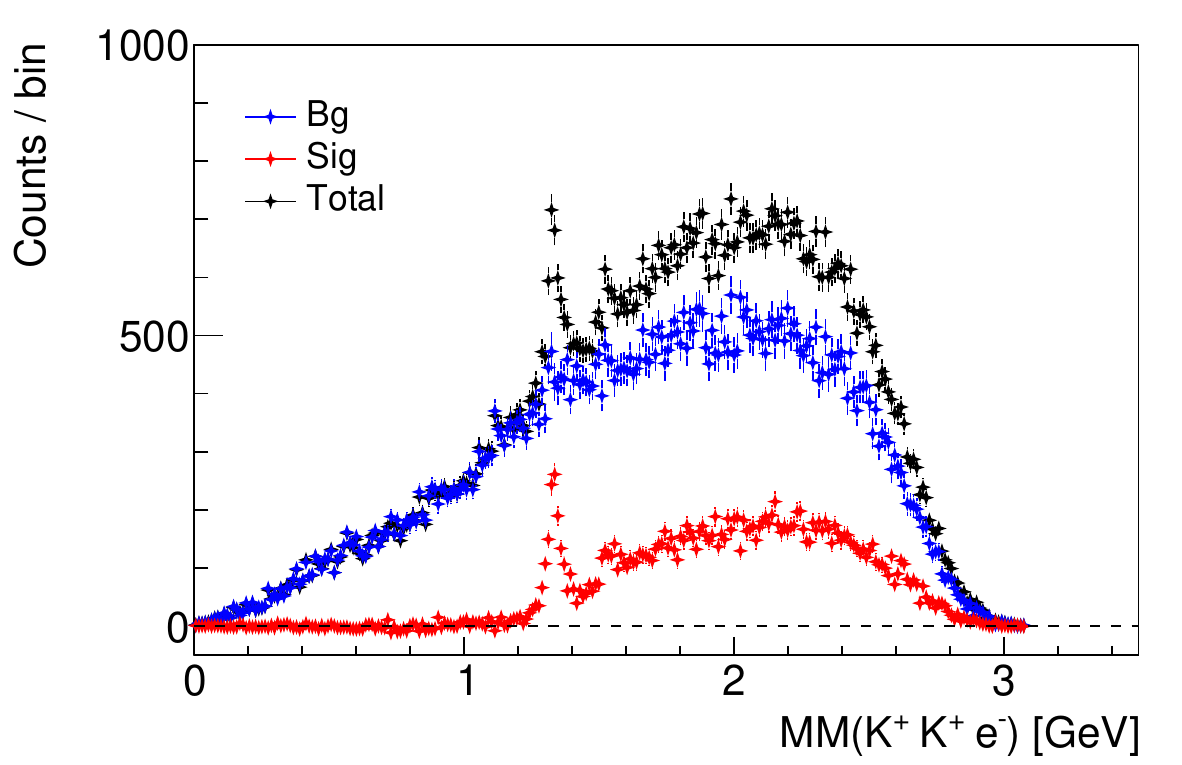}
\includegraphics[width=0.42\textwidth,angle=0]{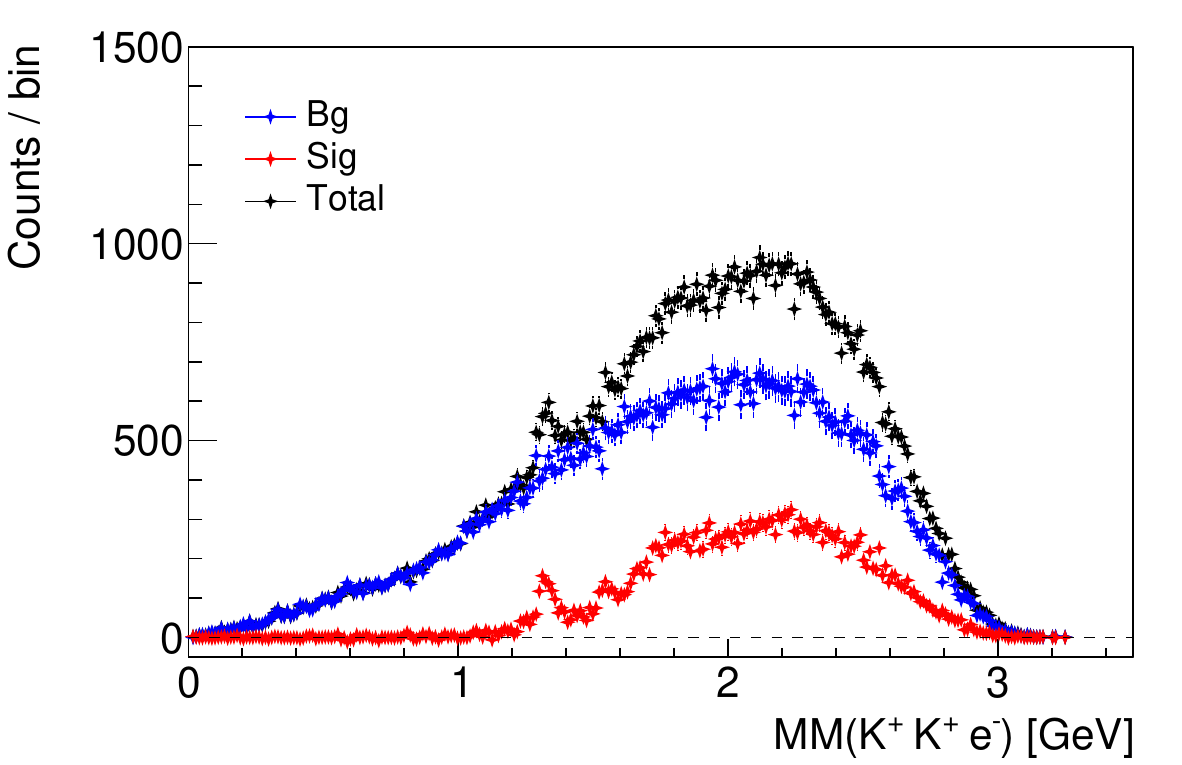}
\includegraphics[width=0.42\textwidth,angle=0]{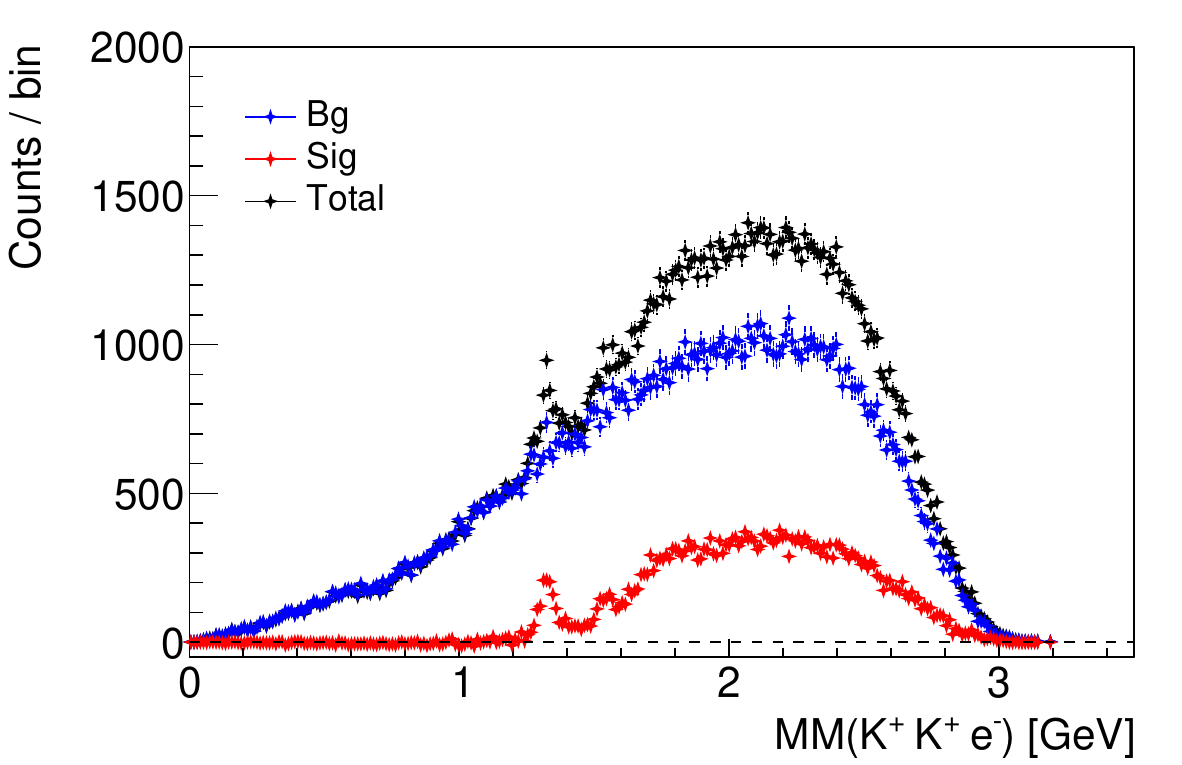}
\end{center}
\caption{Final sWeighted distributions for all of the run periods. Total in blue, signal in pink and background in green. \underline{Top}: Data taken in  Fall 2018  with the torus magnet in the \emph{outbending} configuration FD, Middle: Data taken in Fall 2018  with the torus magnet in the \emph{inbending} configuration FD, \underline{Bottom}: Data taken in Spring 2019  with the torus magnet in the \emph{inbending} configuration FD.}
\label{FinalPlotsFD}
\end{figure}

\begin{figure}[h!]
\begin{center}
\includegraphics[width=0.42\textwidth,angle=0]{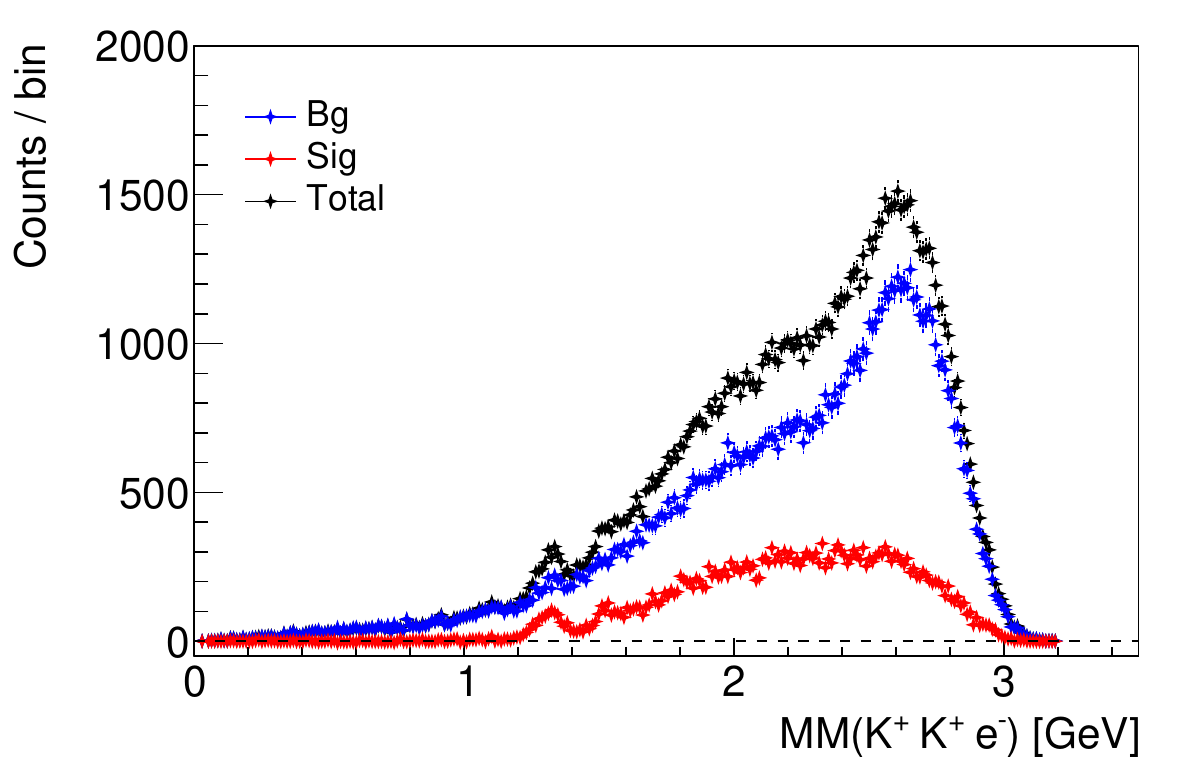}
\includegraphics[
width=0.42\textwidth]{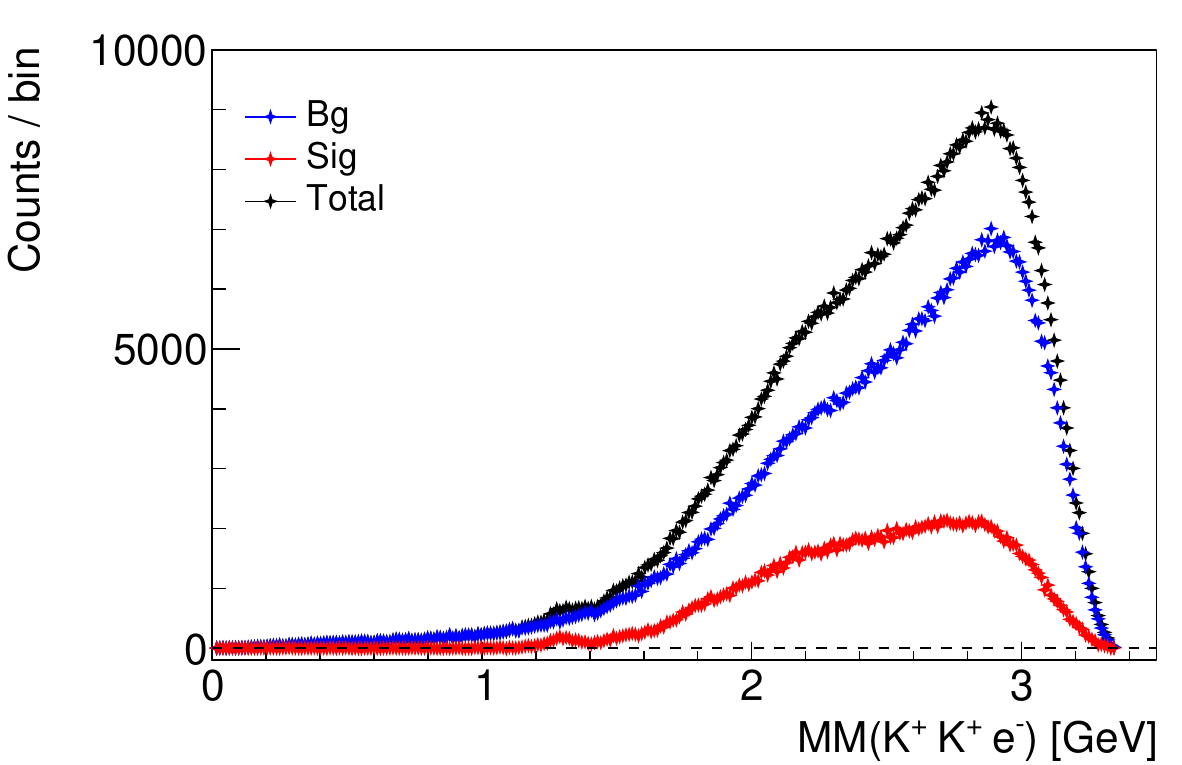}
\includegraphics[width=0.42\textwidth,angle=0]{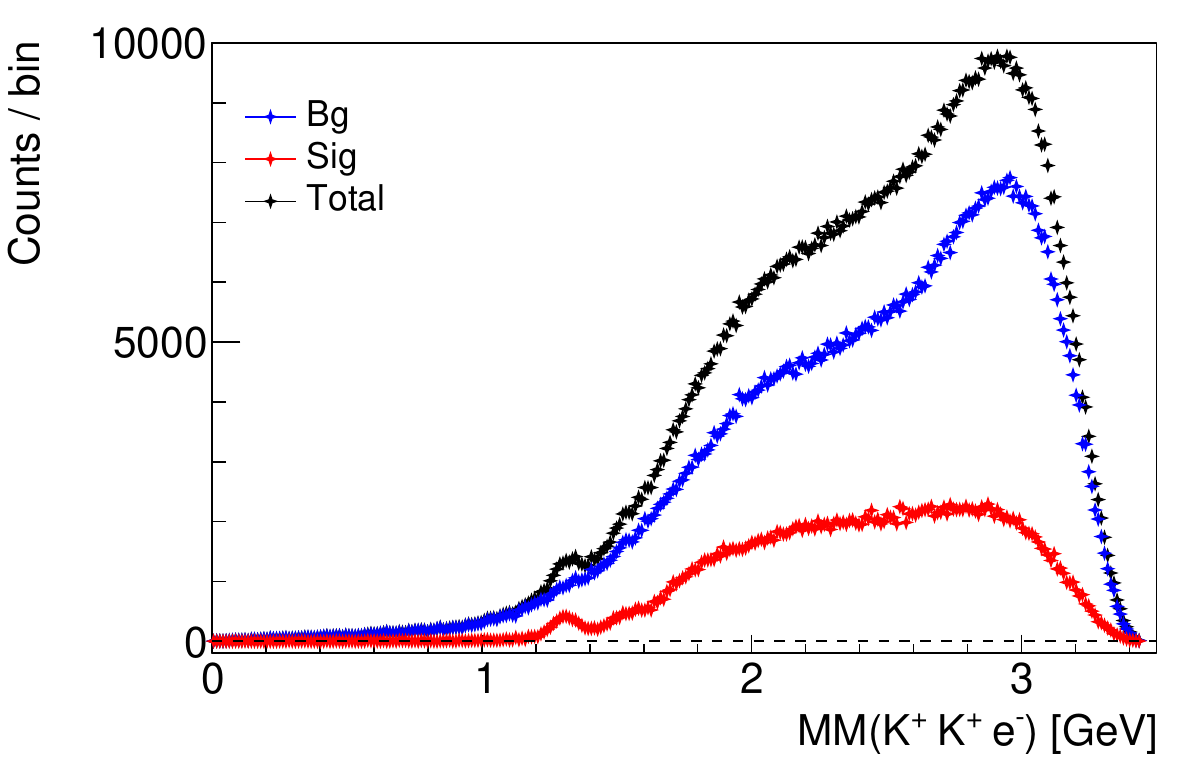}
\end{center}
\caption{Final sWeighted distributions for all of the run periods. Total in blue, signal in pink and background in green. \underline{Top}: Data taken in Fall 2018  with the torus magnet in the \emph{outbending} configuration FT, Middle: Data taken in Fall 2018  with the torus magnet in the \emph{inbending} configuration FT, \underline{Bottom}: Data taken in Spring 2019 with the torus magnet in the \emph{inbending} configuration FT.}
\label{FinalPlotsFT}
\end{figure}

The total, signal and background distributions for each dataset, separated for FD and FT are given in Fig.~\ref{FinalPlotsFD} and Fig.~\ref{FinalPlotsFT} respectively. It is clear that the signal leakage has decreased immensely after the implementation of pion and proton background functions in the total fit function. The consistency of the results across different run periods indicates that a kaon identification method that is robust under different detector and beam conditions has been achieved. 

\end{document}